\documentclass[twocolumn]{svjour3}
\usepackage[american]{babel}
\usepackage[T1]{fontenc}
\usepackage[utf8]{inputenc}
\usepackage{graphicx}
\usepackage{color, colortbl}
\usepackage{enumitem}
\usepackage{comment}
\usepackage{booktabs}
\usepackage{multirow}
\usepackage{listings}
\usepackage{footmisc}
\usepackage{xcolor}
\usepackage{academicons}

\usepackage{array}
\newcolumntype{L}[1]{>{\raggedright\arraybackslash}m{#1}}
\newcolumntype{C}[1]{>{\centering\arraybackslash}m{#1}}
\newcolumntype{R}[1]{>{\raggedleft\arraybackslash}m{#1}}

\usepackage[hyphens]{url}
\newcommand{\furl}[1]{%
	\footnote{\url{#1}}%
}

\usepackage{fmtcount}
\usepackage{numprint}
\npthousandsep{,}
\npdecimalsign{.}
\newcommand{\printNo}[1]{%
	\ifnum #1 > -1
	\ifnum #1 < 11
	\numberstringnum{#1}%
	\else
	\numprint{#1}%
	\fi
	\else
	\numprint{#1}%
	\fi%
}

\definecolor{Gray}{RGB}{242,242,242}
\definecolor{default-lst-background}{RGB}{242,242,242}
\definecolor{default-lst-keyword}{RGB}{94,20,64}
\definecolor{default-lst-string}{RGB}{15,26,250}
\definecolor{default-lst-comment}{RGB}{31,97,46}
\definecolor{default-lst-annotation}{HTML}{6e2c00}
\definecolor{eclipse-java-background}{RGB}{235,235,235}
\definecolor{eclipse-java-keyword}{RGB}{127,0,85}
\definecolor{eclipse-java-string}{RGB}{42,0,255}
\definecolor{eclipse-java-comment}{RGB}{63,127,95}
\definecolor{eclipse-java-annotation}{RGB}{127,159,191}
\definecolor{orcidlogocol}{HTML}{A6CE39}

\lstdefinestyle{lemma}{	
	morekeywords={list, string, structure, import, datatypes, from, as, functional, microservice, microservices, @endpoints, protocols, sync, data, formats, default, with, format, async, public, interface, out, deployment, technologies, operation, environments, container, technology, environment, deploys, values, basic, endpoints, @sync, @async, int, double, required, long, version, context, float, date, in, function, procedure, immutable, hide, types, boolean, primitive, type, byte, short, char, based, on, service, aspects, for, fields, parameters, selector, aspect, true, false, internal, unspecified, operations, used, by, depends, nodes, properties, is},	
	morecomment=[s]{/*}{*/},
	morecomment=[l]{//},
	moredelim={[is][\textcolor{default-lst-annotation}]{\%\%}{\%\%}},
	moredelim={[is][\bfseries\textcolor{lstkeyword}]{§§}{§§}},
	moredelim={[is][\itshape\textcolor{lstfeature}]{\#}{\#}},
	moredelim={[is][\textcolor{default-lst-comment}]{\#\#}{\#\#}},
	moredelim={[is][\itshape\textcolor{default-lst-parameter}]{||}{||}},
	moredelim={[is][\bfseries\textcolor{lsttasktag}]{§}{§}},
	morestring=[b]",
	morestring=[b]'
}

\lstdefinestyle{dockerfile}{	
	morekeywords={RUN, COPY, FROM, CMD, ARG, ENTRYPOINT, EXPOSE},	
	morecomment=[l]{//},
	moredelim={[is][\textcolor{default-lst-annotation}]{\%\%}{\%\%}},
	moredelim={[is][\bfseries\textcolor{lstkeyword}]{§§}{§§}},
	moredelim={[is][\itshape\textcolor{lstfeature}]{\#}{\#}},
	moredelim={[is][\textcolor{default-lst-comment}]{\#\#}{\#\#}},
	moredelim={[is][\itshape\textcolor{lstapiparameter}]{||}{||}},
	moredelim={[is][\bfseries\textcolor{lsttasktag}]{§}{§}},
	morestring=[b]",
	morestring=[b]'
}

\lstdefinestyle{kubernetes}{	
	morekeywords={apiVersion, kind, metadata, creationTimestamp, labels, app, name, spec, replicas, selector, match:Labels, status, ports, name, port, targetPort, status},	
	morecomment=[s]{/*}{*/},
	morecomment=[l]{//},
	moredelim={[is][\textcolor{default-lst-annotation}]{\%\%}{\%\%}},
	moredelim={[is][\bfseries\textcolor{lstkeyword}]{§§}{§§}},
	moredelim={[is][\itshape\textcolor{lstfeature}]{\#}{\#}},
	moredelim={[is][\textcolor{lstapicomment}]{\#\#}{\#\#}},
	moredelim={[is][\itshape\textcolor{lstapiparameter}]{||}{||}},
	moredelim={[is][\bfseries\textcolor{lsttasktag}]{§}{§}},
	morestring=[b]",
	morestring=[b]'
}

\lstdefinestyle{property}{	
	morekeywords={spring, application, name, server, port, eureka, instance, preferIpAddress, client, registerWithEureka, fetchRegistry, serviceUrl, defaultZone},	
	morecomment=[s]{/*}{*/},
	moredelim={[is][\textcolor{default-lst-annotation}]{\%\%}{\%\%}},
	moredelim={[is][\bfseries\textcolor{lstkeyword}]{§§}{§§}},
	moredelim={[is][\itshape\textcolor{lstfeature}]{\#}{\#}},
	moredelim={[is][\textcolor{lstapicomment}]{\#\#}{\#\#}},
	moredelim={[is][\itshape\textcolor{lstapiparameter}]{||}{||}},
	moredelim={[is][\bfseries\textcolor{lsttasktag}]{§}{§}},
	morestring=[b]",
	morestring=[b]'
}

\lstdefinelanguage{JSON}{
    string=[s]{"}{"},
    stringstyle=\color{default-lst-keyword},
    comment=[l]{:},
    commentstyle=\color{black},
    moredelim={[is][\itshape\textcolor{default-lst-comment}]{/*}{*/}}
}

\usepackage[hidelinks]{hyperref}
\usepackage[capitalise]{cleveref}
\crefformat{section}{Sect.~#2#1#3}
\Crefformat{section}{Section~#2#1#3}
\crefname{subsection}{Subsect.}{Subsects.}
\crefformat{subsection}{Subsect.~#2#1#3}
\crefname{subsubsection}{Subsubsect.}{Subsubsects.}
\Crefformat{subsubsection}{Subsubsection~#2#1#3}
\crefformat{subsubsection}{Subsubsect.~#2#1#3}
\crefformat{figure}{Fig.~#2#1#3}
\Crefformat{figure}{Figure~#2#1#3}
\crefname{lstlisting}{Listing}{Listings}
\Crefname{lstlisting}{Listing}{Listings}

\def \zvalLocDeploymentSpecs{285}
\def \zvalLocGenModelsCustomerCore{171}
\def \zvalLocGenModelsCustomerManagement{174}
\def \zvalLocManModelsLm{3702}
\def \zvalLocManModelsLmOperation{311}
\def \zvalLocManModelsCustomerCore{588}
\def \zvalLocManModelsCustomerManagement{458}
\def \zvalLocOpenApiCustomerCoreJSON{534}
\def \zvalLocOpenApiCustomerManagementJSON{496}
\def \zvalLocManModelsCustomerCoreOperationLEMMA{50}
\def \zvalLocManModelsCustomerManagementOperationLEMMA{49}

\makeatletter
\def\cl@chapter{\@elt {theorem}}
\makeatother

\begin{document}
\sloppy
\title{Applying Model-Driven Engineering to Stimulate the Adoption of DevOps Processes in Small and Medium-Sized Development Organizations\thanks{Attention. This is a preprint draft version.}}
\subtitle{The Case for Microservice Architecture}
\titlerunning{MDE to Stimulate DevOps Processes in the Context of Microservice Architecture}

\author{
    Jonas Sorgalla* \and
    Philip Wizenty \and
	Florian Rademacher \and
	Sabine Sachweh \and
	Albert Z\"undorf
}

\authorrunning{Sorgalla et al.}

\institute{J. Sorgalla* \at
              IDiAL Institute, University of Applied Sciences and Arts Dortmund, Otto-Hahn-Stra\ss{}e 27, 44227 Dortmund, Germany\\
              \email{jonas.sorgalla@fh-dortmund.de}\\
              ORCiD: 0000-0002-7532-7767
           \and
           P. Wizenty \at
              IDiAL Institute, University of Applied Sciences and Arts Dortmund, Otto-Hahn-Stra\ss{}e 27, 44227 Dortmund, Germany\\
              \email{philip.wizenty@fh-dortmund.de}\\
              ORCiD: 0000-0002-3588-5174
            \and
            F. Rademacher \at
              IDiAL Institute, University of Applied Sciences and Arts Dortmund, Otto-Hahn-Stra\ss{}e 27, 44227 Dortmund, Germany\\
              \email{florian.rademacher@fh-dortmund.de}\\
              ORCiD: 0000-0003-0784-9245
            \and
            S. Sachweh \at
              IDiAL Institute, University of Applied Sciences and Arts Dortmund, Otto-Hahn-Stra\ss{}e 27, 44227 Dortmund, Germany\\
              \email{sabine.sachweh@fh-dortmund.de}
            \and
            A. Zündorf \at
              Department of Computer Science and Electrical Engineering, University of Kassel, Wilhelmsh\"oher Allee 73, 34121 Kassel, Germany\\
              \email{zuendorf@uni-kassel.de}  
}

\date{Received: date / Accepted: date}

\maketitle


\begin{abstract}
\phantom{forceLineBreak}

\noindent\textit{Purpose:} Microservice Architecture (MSA) denotes an increasingly popular architectural style in which business capabilities are wrapped into autonomously de\-vel\-opable and de\-ploy\-able software components called microservices. Microservice applications are developed by multiple DevOps teams each owning one or more services. In this article, we explore the state of how DevOps teams in small and medium-sized organizations (SMOs) cope with MSA and how they can be supported.

\noindent\textit{Methods:} We show through a secondary analysis of an exploratory interview study comprising six cases, that the organizational and technological complexity resulting from MSA poses particular challenges for small and medium-sized organizations (SMOs). We apply Model-Driven Engineering to address these challenges. 

\noindent\textit{Results:} As results of the second analysis, we identify the challenge areas of building and maintaining a common architectural understanding, and dealing with deployment technologies. To support DevOps teams of SMOs in coping with these challenges, we present a model-driven workflow based on LEMMA---the Language Ecosystem for Modeling Microservice Architecture. To implement the workflow, we extend LEMMA with the functionality to (i) generate models from API documentation; (ii) reference remote models owned by other teams; (iii) generate deployment specifications; and (iv) generate a visual representation of the overall architecture. 

\noindent\textit{Conclusion:} We validate the model-driven workflow and our extensions to LEMMA through a case study showing that the added functionality to LEMMA can bring efficiency gains for DevOps teams. To develop best practices for applying our workflow to maximize efficiency in SMOs, we plan to conduct more empirical research in the field in the future.

\keywords{DevOps, Microservice Architecture, Development Process, Model-driven Engineering}
\end{abstract}

\section{Introduction}
Microservice Architecture (MSA) is a novel architectural style for service-based software systems with a strong focus on loose functional, technical, and organizational coupling of services~\cite{Newman2015}. In a microservice architecture, services are tailored to distinct business capabilities and executed as independent processes. The adoption of MSA is expected to increase an application's scalability, maintainability, and reliability~\cite{Newman2015}. It is frequently employed to decompose monolithic applications for which such quality attributes are of critical scale~\cite{Bogner2019}.

MSA fosters the adoption of DevOps practices, because it promotes to (i) bundle microservices in self-contained deployment units for continuous delivery; and (ii) delegate responsibility for a microservice to a single team being composed of members with heterogeneous professional backgrounds~\cite{Balalaie2016b,Nadareishvili2016}. Conway's Law~\cite{Conway1968} is a determining factor in DevOps-based MSA engineering. It states that the communication structure of a system reflects the structure of its development organization. Thus, in order to achieve loose coupling and autonomy of microservices, it is also crucial to divide the responsibility for microservices' development and deployment between autonomous DevOps teams~\cite{Nadareishvili2016}. As a result, MSA engineering leads to a distributed development process, in which several teams create coherent services of the same software system in parallel. 

While various larger enterprises like Netflix\furl{https://netflix.github.io}, Spotify\furl{https://labs.spotify.com}, or Zalando\furl{https://opensource.zalando.com} regularly report about their successful adoption of MSA, there are only a small number of experience reports (e.g.,~\cite{Buchgeher2017}) about how microservices combined with DevOps can be successfully implemented in small and medium-sized development organizations (SMOs) with less than 100 developers involved. Such SMOs typically do not have sufficient resources to directly apply large-scale process models such as Scrum at Scale~\cite{Conboy2019,ScrumScale2020} in terms of employees, knowledge, and experience.

To support SMOs in bridging the gap between available resources and required effort for a successful adoption of DevOps-based MSA engineering, we (i) investigate the characteristics of small- and medium-scale microservice development processes; and (ii) propose means to reduce complexity and increase productivity in DevOps-based MSA engineering within SMOs. More precisely, the contributions of our article are threefold. First, we identify challenges of SMOs in DevOps-based MSA engineering by analyzing a data set of an exploratory qualitative study and linking it with existing empirical knowledge. Second, we employ Model-driven Engineering (MDE)~\cite{Combemale2017} to introduce a workflow for coping with the previously identified challenges in DevOps-based MSA engineering for SMOs. Third, we present and validate extensions to LEMMA (Language Ecosystem for Modeling Microservice Architecture), which is a set of Eclipse-based modeling languages and model transformations for MSA engineering~\cite{Rademacher2020} enabling sophisticated modeling support for the workflow.

The remainder of this article is organized as follows. In \cref{sec:msa}, we describe in detail the microservice architecture style particularly related to the design, development, and operation stages. In addition, we explain organizational aspects that result from the use of microservices. Section~\ref{sec:lemma} illustrates LEMMA as a set of modeling languages and tools that address the MDE of MSA. In \cref{sec:qualitative-study}, we analyze a dataset based on an exploratory interview study in SMOs to identify challenging areas in engineering MSA for DevOps teams in SMOs. Based on these challenge areas, we present a model-driven workflow in \cref{sec:workflow} and describe the extensions of LEMMA in order to support the workflow. In this regard, \cref{subsec:apispecifications} present means to derive LEMMA models from API documentation, \cref{subsec:assemblingSystemModel} presents extensions to the LEMMA languages in order to assemble individual microservice models, \cref{subsec:lemmaVisualizer} describes additions to create a visual representation of microservice models, \cref{subsec:InfrastructureModels} presents means to specify deployment infrastructure, and \cref{subsec:InfrastructureCodeGeneration} elaborates on the ability to generate infrastructure code. We validate our contributions to LEMMA  in \cref{sec:validation}. Section~\ref{sec:discussion} discusses the model-driven workflow and LEMMA components towards the application in DevOps teams of SMOs. We present related research in \cref{sec:related-work}. The article ends with a conclusion and outlook on future work in \cref{sec:conclusion}.

\section{Background}
\label{sec:msa}
This section provides background on the MSA approach and its relation towards the DevOps paradigm. It details special characteristics in the design, development, operation, and organization of microservice architectures and their realization.

\subsection{General}\label{subsec:MSAGeneral}
MSA is a novel approach towards the design, development, and operation of service-based software systems~\cite{Newman2015}. Therefore, MSA promotes to decompose the architecture of complex software systems into \emph{services}, i.e., loosely coupled software components that interact by means of predefined interfaces and are composable to synergistically realize coarse-grained business logic~\cite{Erl2005}.

Compared to other approaches for ar\-chi\-tect\-ing ser\-vice-based software systems, e.g., SOA~\cite{Erl2005}, MSA puts a strong emphasis on \emph{service-specific independence}. This independence distinguishes MSA from other approaches w.r.t. the following features~\cite{Newman2015,Nadareishvili2016,Richards2016}:
\begin{itemize}
    \item Each microservice in a microservice architectures focuses on the provisioning of a single distinct capability for functional or infrastructure purposes.
    \item A microservice is independent from all other architecture components regarding its implementation, data management, testing, deployment, and operation.
    \item A microservice is fully responsible for all aspects related to its interaction with other architecture components, ranging from the determination of communication protocols over data and format conversions to failure handling.
    \item Exactly one team is responsible for a microservice and has full accountability for its services' design, development, and deployment.
\end{itemize}

Starting from the above features, the adoption of MSA may introduce increases in quality attributes~\cite{ISO25010} such as (i) scalability, as it is possible to purposefully run new instances of microservices covering strongly demanded functionality; (ii) maintainability, as microservices are seamlessly replaceable with alternative implementations; and (iii) reliability, as it delegates responsibility for robustness and resilience to microservices~\cite{Newman2015,Dragoni2017,Dragoni2018}. Additionally, MSA fosters DevOps and agile development, because its single-team ownership calls for heterogeneous team composition and microservices' constrained scope fosters their evolvability~\cite{Taibi2019,Francesco2017}.

Despite its potential for positively impacting the aforementioned features of a software architecture and its implementation, MSA also introduces complexity both to development processes and operation~\cite{Francesco2017,Taibi2017,Soldani2018}. Consequently, practitioners in SMOs perceive the successful adoption of MSA as complex~\cite{Bogner2019}. Challenges that must be addressed in MSA adoption are spread across all stages in the engineering process, and thus concern the design of the architecture, its development and operation. Furthermore, MSA imposes additional demands on the organization of the engineering process.

\subsection{Design Stage}
A frequent design challenge in MSA engineering concerns the decomposition of an application domain into microservices that each have a suitable \emph{functional granularity}~\cite{Haesen2008,Soldani2018}. Too coarse-grained microservice capabilities neglect the aforementioned benefits of MSA in terms of service-specific independence. Too fine-grained microservices, on the other hand, may require an inefficiently high amount of communication and thus network traffic at runtime~\cite{Kratzke2017b}. Although there exist approaches such as Domain-driven Design (DDD)~\cite{Evans2004} to support in the systematic decomposition and granularity determination of a microservice architecture~\cite{Newman2015}, their perceived complexity hampers widespread adoption in practice~\cite{Francesco2018,Bogner2019}.

An additional specific in microservice design stems from MSA's omission of explicit \emph{service contracts}~\cite{Papazoglou2008}. By contrast to SOA, MSA considers the API of a microservice its implicit contract~\cite{Zimmermann2017}, thereby delegating concerns in API management, e.g., API versioning to microservices~\cite{Soldani2018}. Consequently, microservices must ensure their compatibility with possible consumers and also inform them about possible interaction requirements. Furthermore, implicit microservice contracts foster ad hoc communication, which increases runtime complexity and the occurrence of cyclic interaction relationships~\cite{Taibi2018}.

\subsection{Development Stage}\label{subsec:DeploymentStage}
By contrast to monolithic applications, which rely on a holistic, yet vendor-dependent technology stack~\cite{Dragoni2017}, microservice architectures foster \emph{technology heterogeneity}~\cite{Newman2015}. Specifically, due to the increase in service-specific independence, each microservice may employ those technologies that best fit a certain capability. Typical \emph{technology variation points}~\cite{Rademacher2019} comprise programming languages, databases, communication protocols, and data formats. However, technology heterogeneity imposes a greater risk for technical debt, additional maintainability costs, and steeper learning curves, particularly for new members of a microservice team~\cite{Taibi2018}.

\subsection{Operation Stage}\label{subsec:OperationStage}
MSA usually requires a sophisticated deployment and operation infrastructure consisting of, e.g., continuous delivery systems, a basic container technology and orchestration platform, to cope with MSA's emphasis of maintainability and reliability~\cite{Taibi2019}. In addition, microservices often rely on further infrastructure components such as service discoveries, API gateways, or monitoring solutions~\cite{Balalaie2016}, which lead to additional administration and maintenance effort. Consequently, microservice operation involves a variety of different technical components, thereby resulting in a significant complexity increase compared to monolithic applications~\cite{Soldani2018}.

Furthermore, technology heterogeneity also concerns microservice operation w.r.t. technology variation points like deployment and infrastructure technologies~\cite{Rademacher2019}. Particularly the latter also involve independent decision-making by microservice teams. For example, there exist infrastructure technologies, e.g., to increase performance or resilience, which directly focus on a microservice~\cite{Balalaie2016}. Hence, teams are basically free to decide for suitable solutions based on different criteria such as compatibility with existing microservice implementations or available experience.

\subsection{Organizational Aspects}
\label{subsec:MicroserviceDevProcess}

The use of MSA requires a compatible organizational structure, i.e., following Conway's law, a structure that corresponds to the communication principle of microservices. This results in the necessity of using separate teams, each of which is fully responsible for one or more services (cf.~\cref{subsec:MSAGeneral}). 
The requirement that a team should cover the entire software lifecycle of its microservices automatically leads to the need for cross-functional teams. In order to ensure collaboration between teams, large companies such as Netflix or Spotify usually use established \emph{large-scale agile} process models~\cite{Dikert2016}, e.g., the Scaled Agile Framework (SAFe)~\cite{SAFe2019}, Scrum at Scale~\cite{ScrumScale2020}, or the Spotify Model~\cite{Smite2019}. Establishing such a form of organization and to establish \emph{organizational alignment} may require upfront efforts~\cite{Newman2015}. 

Thus, MSA fosters DevOps practices, which can result in lowered cost and accelerate the pace of product increments~\cite{Nadareishvili2016}. To this end, it is critical to foster a \emph{collaborative culture} within and across teams to promote integration and collaboration among team members with different professional backgrounds~\cite{Luz2018}.

A key enabler of a collaborative culture is the extensive \emph{automation} of manual tasks to prevent the manifestation of inter-team and extra-team silos~\cite{Luz2018}. Specifically, it relieves people from personal accountability for a task and may thus help in reducing existing animosities of team members with different professional backgrounds~\cite{Knoche2019}.

Another pillar of a collaborative culture is \emph{knowledge sharing} following established formats and guidelines~\cite{Luz2018}. It aims to mitigate the occurrence of insufficient communication, which can be an impediment in both MSA and DevOps~\cite{Francesco2017,Riungu2016}.

\section{Language Ecosystem for Modeling Microservice Architecture}
\label{sec:lemma}
In our previous works we developed LEMMA~\cite{Rademacher2019,Rademacher2020}. LEMMA is a set of Eclipse-based modeling languages and model transformations that aims to mitigate the challenges in MSA engineering (cf.~\cref{sec:msa}) by means of Model-driven Engineering (MDE)~\cite{Combemale2017}.

To this end, LEMMA refers to the notion of \emph{architecture viewpoint}~\cite{ISO42010} to support stakeholders in MSA engineering in organizing and expressing their concerns towards a microservice architecture under development. More specifically, LEMMA clusters four viewpoints on microservice architectures. Each viewpoint targets at least one stakeholder group in MSA engineering, and comprises one or more stakeholder-oriented modeling languages.

The modeling languages enable the construction of microservice architecture models and their composition by means of an import mechanism. As a result, LEMMA allows reasoning about coherent parts of a microservice architecture~\cite{ISO42010}, e.g., to assess quality attributes and technical debt of microservices~\cite{Rademacher2020b} or perform DevOps-oriented code generation~\cite{Rademacher2020c}.

The following paragraphs summarize LEMMA's approach to microservice architecture model construction and processing.

\subsection{Microservice Architecture Model Construction}\label{subsec:MicroserviceArchitectureModelConstruction}
\begin{figure*}
    \centering
    \includegraphics[scale=0.4]{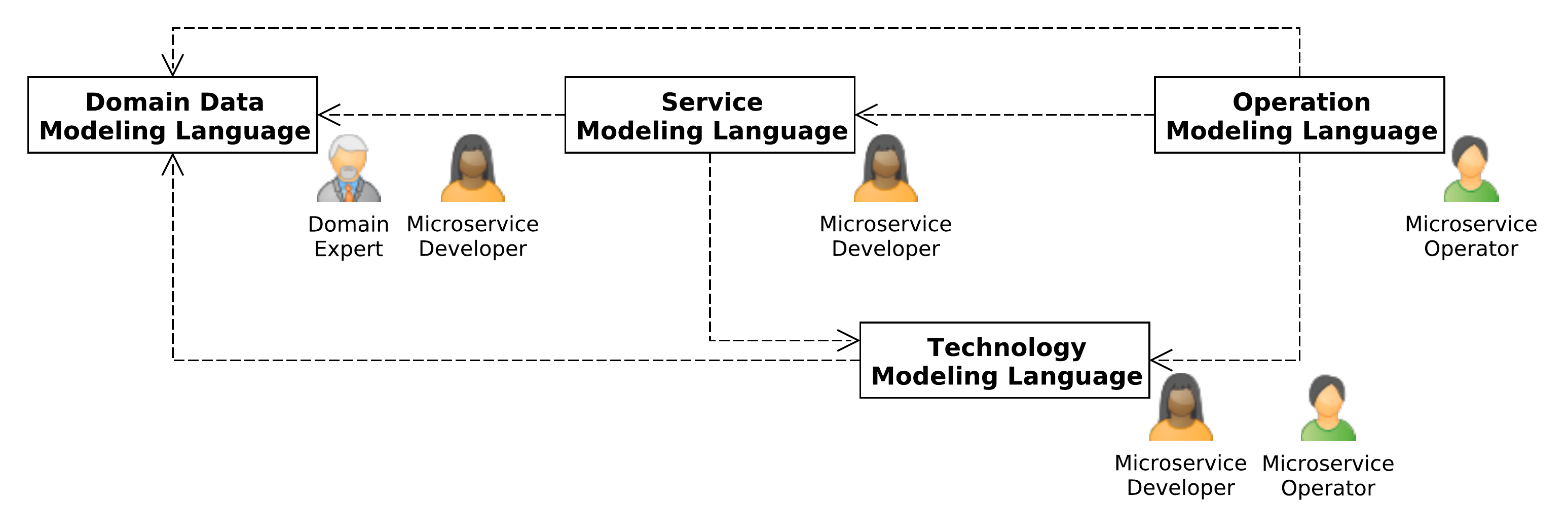}
    \caption{Overview of LEMMA's modeling languages, their compositional dependencies and addressed stakeholders. Arrow semantics follow those of UML for dependency specifications~\cite{OMG_UML_2017}.}
    \label{fig:lemma-languages}
\end{figure*}

\Cref{fig:lemma-languages} provides an overview of LEMMA's modeling languages, their compositional dependencies and the addressed stakeholders in MSA engineering.

LEMMA's Domain Data Modeling Language~\cite{Rademacher2020} allows model construction in the context of the \emph{domain viewpoint} on a microservice architecture. Therefore, it addresses the concerns of domain experts and microservice developers. First, the language aims to mitigate the complexity of DDD (cf.~\cref{sec:msa}) by defining a minimal set of modeling concepts for the construction of domain concepts, i.e., data structures and list types, and the assignment of DDD patterns, e.g., Entity or Value Object~\cite{Evans2004}. Additionally, it integrates  validations to ensure the semantically correct usage of the patterns. Second, the language considers underspecification in DDD-based domain model construction~\cite{Rademacher2018b}, thereby facilitating model construction for domain experts. However, microservice developers may later resolve underspecification to enable automated model processing~\cite{Rademacher2020c}. All other LEMMA modeling languages depend on the Domain Data Modeling Language (cf.~\cref{fig:lemma-languages}) because it provides them with a Java-aligned type system~\cite{Rademacher2020} given Java's predominance in service programming~\cite{Schermann2016,Bogner2019}.

LEMMA's Service Modeling Language~\cite{Rademacher2020} addresses the concerns of microservice developers (cf.~\cref{fig:lemma-languages}) in the \emph{service viewpoint} on a microservice architecture. One goal of the Service Modeling Language is to make the APIs of microservices explicit (cf.~\cref{sec:msa}) but keeping their definition as concise as possible based on built-in language primitives. That is, the language provides developers with targeted modeling concepts for the definition of microservices, their interfaces, operations and endpoints. LEMMA service models may import LEMMA domain models to identify the responsibility of a microservice for a certain portion of the application domain~\cite{Newman2015} and type operation parameters with domain concepts.

LEMMA's Technology Modeling Language~\cite{Rademacher2019} considers technology to constitute a dedicated architecture viewpoint~\cite{Hofmeister2007} that frames the concerns of tech\-nol\-o\-gy-savvy stakeholders in MSA engineering, i.e., microservice developers and operators (cf.~\cref{fig:lemma-languages}). The Technology Modeling Language enables those stakeholder groups to construct and apply technology models. A LEMMA technology model modularizes information targeting a certain technology relevant to microservice development and operation, e.g., programming languages, software frameworks, or deployment technologies. Furthermore, it integrates a generic metadata mechanism based on \emph{technology aspects}~\cite{Rademacher2019}. Technology aspects may, for example, cover annotations of software frameworks. LEMMA service and operation models depend on LEMMA technology models (cf.~\cref{fig:lemma-languages}) and import them to apply the contained technology information to, e.g., modeled microservices and containers. In particular, LEMMA's Technology Modeling Language aims to cope with technology heterogeneity in MSA engineering (cf.~\cref{sec:msa}) by making technology decisions explicit~\cite{Soliman2015}.

LEMMA's Operation Modeling Language~\cite{Rademacher2020} addresses the concerns of microservice operators (cf.~\cref{fig:lemma-languages}) w.r.t. the \emph{operation viewpoint} in MSA engineering. To this end, the language integrates primitives for the concise modeling of microservice containers, infrastructure nodes, and technology-specific configuration. To model the deployment of microservices, LEMMA operation models import LEMMA service models and assign modeled microservices to containers. Additionally, it is possible to express the dependency of containers on infrastructure nodes such as service discoveries or API gateways~\cite{Balalaie2016}. By providing microservice operators with a dedicated modeling language we aim to cope with operation challenges in MSA engineering (cf.~\cref{sec:msa}). First, the Operation Modeling Language defines a unified syntax for the modeling of heterogeneous operation nodes of a microservice architecture. Second, it is flexibly extensible with support for operation technologies, e.g., for microservice monitoring or security, leveraging LEMMA technology models (cf.~\cref{fig:lemma-languages}). Third, operation models may import other operation models, e.g., to compose the models of different microservice teams to centralize specification and maintenance of shared infrastructure components such as service discoveries and API gateways.

\subsection{Microservice Architecture Model Processing}
\label{subsubsec:intermediatemodelprocessing}
LEMMA relies on the notion of \emph{intermediate model representation}~\cite{Jezequel2012} to facilitate the processing of constructed models. To this end, LEMMA integrates a set of \emph{intermediate metamodels} and \emph{intermediate model transformations}. The intermediate metamodels define the concepts to which the elements of an intermediate model for a LEMMA model conform. An intermediate model transformation is then responsible for the automated derivation of an intermediate model from a given input LEMMA model.

This approach to model processing yields several benefits. First, intermediate metamodels decouple modeling languages from model processors. Consequently, languages can evolve independently from processors as long as the intermediate metamodels remain stable. For example, it becomes possible to introduce syntactic sugar in the form of additional shorthand notations for language constructs. Second, intermediate metamodels enable to incorporate language semantics into intermediate models so that model processors need not anticipate them. For instance, LEMMA allows modeling of default protocols for communication types within technology models. In case a service model does not explicitly determine a protocol, e.g., for a microservice, the default protocol of the service's technology model applies implicitly. The intermediate transformation, which converts a service model into its intermediate representation, makes the default protocol explicit. Thus, model processors can directly rely on this information and need not determine the effective protocol for a microservice themselves.

Next to intermediate model representations, LEMMA also provides a model processing framework\furl{https://github.com/SeelabFhdo/lemma/tree/master/de.fhdo.lemma.model_processing}, which facilitates the implementation of Java-based model processors, e.g., for microservice developers without a strong background in MDE. To this end, the framework leverages the Inversion of Control (IoC) design approach~\cite{Johnson1988}, and its realization based on the Abstract Class pattern~\cite{Sobernig2010} and Java annotations~\cite{ORACLE_JAVA_SE13_2019}. In addition, the framework implements the Phased Construction model transformation design pattern~\cite{Lano2014}. That is, the framework consists of several phases including phases for model validation and code generation. To implement a phase as part of a model processor, developers need to provide an implementation of a corresponding abstract framework class, e.g., \texttt{Abstract\-Code\-Generation\-Module}, and augment the implementation with a phase-specific annotation, e.g., \texttt{@Code\-Generation\-Module}. At runtime, model processors pass control over the program flow to the framework. The framework will then (i) parse all given intermediate LEMMA models; (ii) transform them into object graphs, which abstract from a concrete modeling technology; and (iii) invoke the processor-specific phase implementations with the object graphs. As a result, the added complexities of MDE w.r.t. model parsing and the construction of Abstract Syntax Trees as instantiations of language metamodels~\cite{Combemale2017} remain opaque for model processor developers. Moreover, LEMMA's model processing framework provides means to develop model processors as standalone executable Java applications. This characteristic is crucial for the integration of model processors into continuous integration pipelines~\cite{Jongeling2019}, which constitute a component in DevOps-based MSA engineering~\cite{Bass2015,Bogner2019}.

\begin{figure*}[ht]
    \centering
    \includegraphics[scale=0.5]{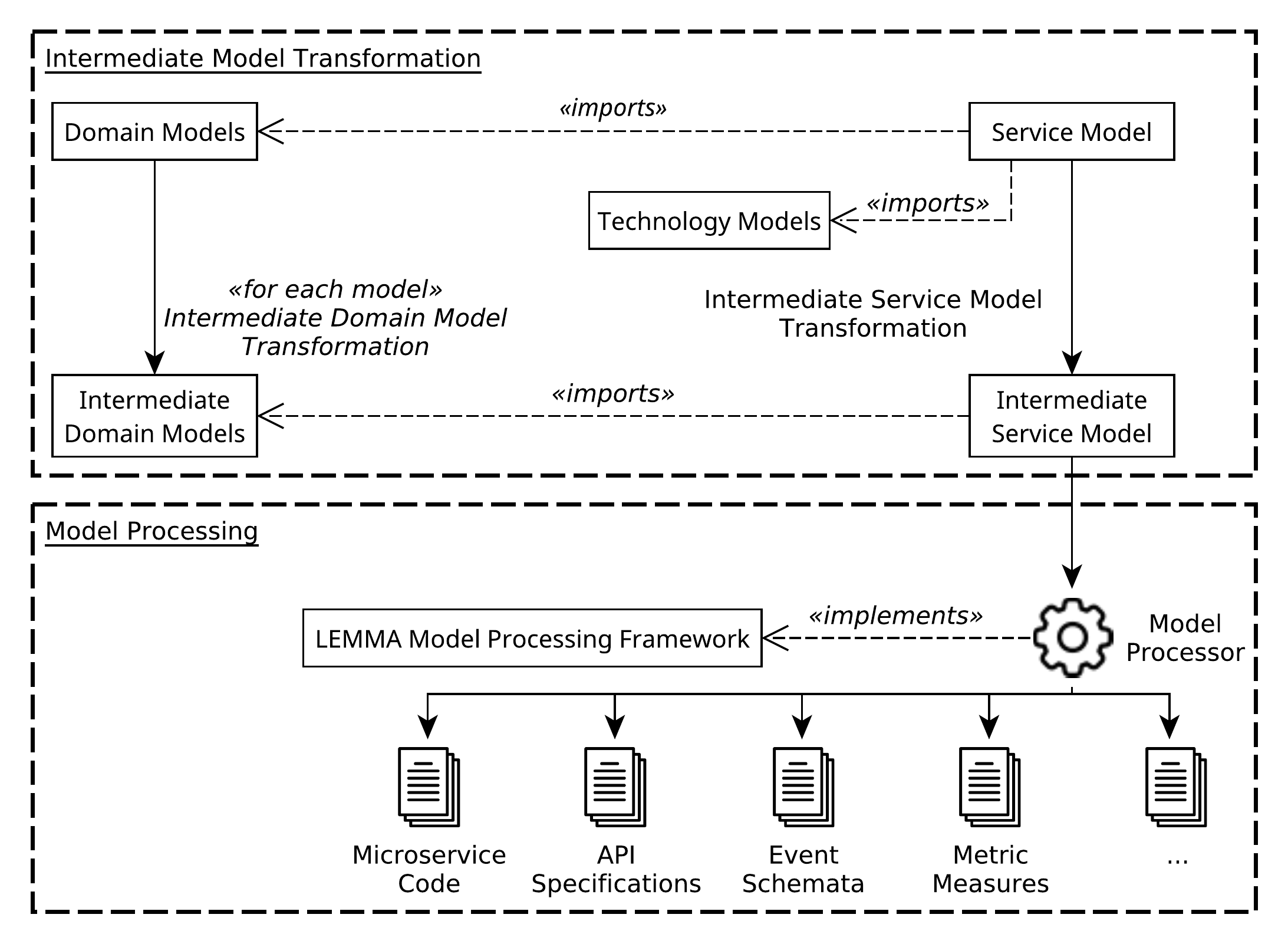}
    \caption{Example for the interplay of LEMMA's intermediate model transformations and model processors based on its model processing framework.}
    \label{fig:model-processing-example}
\end{figure*}

\Cref{fig:model-processing-example} illustrates the interplay of intermediate model transformations, and the implementation and execution of model processors with LEMMA.

\Cref{fig:model-processing-example} comprises two compartments.

The first compartment shows the structure of intermediate model transformations with LEMMA based on the example of a service model constructed with LEMMA's Service Modeling Language. The service model imports a variety of domain models and technology models constructed with LEMMA's Domain Data Modeling Language and Technology Modeling Language, respectively. As a preparatory step, the service model is transformed into its intermediate representation by means of LEMMA's Intermediate Service Model Transformation. Similarly, each imported domain model is transformed into a corresponding intermediate domain model leveraging LEMMA's Intermediate Domain Model Transformation. To this end, the transformation algorithm restores the existing import relationships between service models and domain models for their derived intermediate representations. However, the algorithm does not invoke intermediate transformations on technology models imported by service models. Instead, the applied technology information becomes part of intermediate service models so that model processors can directly access them. Therefore, LEMMA treats technology models and model processors as conceptual unities, i.e., a model processor for a certain technology must be aware of the semantics of the elements in its technology model and be capable in interpreting their application, e.g., within service models.

The second compartment of \cref{fig:model-processing-example} concerns model processing. A LEMMA model processor constitutes an implementation conform to LEMMA's model processing framework, which thus provides the processor with capabilities for model parsing and phase-oriented model processing. Typical results from processing service models comprise (i) executable microservice code; (ii) shareable API specifications, e.g., based on OpenAPI\furl{http://spec.openapis.org/oas/v3.1.0}; (iii) event schemata, e.g., for Apache Avro\furl{https://avro.apache.org}; and (iv) measures of static complexity and cohesion metrics applicable to MSA~\cite{Hirzalla2009,Athanasopoulos2015,Haupt2017,Engel2018,Bogner2020}.

\section{DevOps-Related Challenges in Microservice Architecture Engineering of SMOs}
\label{sec:qualitative-study}

In this section, we present an empirical analysis of microservice development processes (cf.~\cref{subsec:MicroserviceDevProcess}) in SMOs with the goal of identifying SMO-specific challenges in microservice engineering. For this purpose, we perform a \emph{secondary analysis}~\cite{Heaton1998} of transcribed qualitative interviews from one of our previous works~\cite{SorgallaInterviews2020}. While the initial analysis of the data through inductive open coding, our secondary analytical procedure specifically aims to identify challenges and obstacles during the development process.

\subsection{Study Design}
The study from which the dataset emerged is a as a comparative multi-case study~\cite{yin2017}. The aim of the study was to gain exploratory insights into the development processes of SMOs. To this end, in-depth interviews were conducted on-site in 2019 with five software architects, each from a different company, and afterwards transcribed. The interviews were conducted in a semi-structured manner and covered the areas of (i) applied development process; (ii) daily routines; (iii) meeting formats; (iv) tools; (v) documentation; and (vi) knowledge management. Participants were recruited from existing contacts of our research group to SMOs. Furthermore, we constrained participant selection to the professional level or senior software architects, and SMOs that develop microservice systems with equal or less than 100 people. 

\subsection{Dataset}
\label{subsec:interviewcases}
As depicted in \cref{tab:interviewcases}, the dataset includes transcripts and derived paraphrases covering six different cases (Column C) of microservice development processes in SMOs. In total, we conducted five in-depth interviews (Column I) with software architects whereby I4 covered two cases.

\begin{table*}[htp]
	\centering
	\caption{Overview of explored SMO cases~\cite{SorgallaInterviews2020}.}
	\label{tab:interviewcases}
	\begin{tabular}{L{0.05\textwidth} L{0.05\textwidth} L{0.15\textwidth} L{0.21\textwidth} R{0.14\textwidth} R{0.08\textwidth} R{0.14\textwidth} }
		\toprule
		\bfseries C & \bfseries I & \bfseries Type & \bfseries Domain & \bfseries \#Services & \bfseries \#Ppl & \bfseries \#Teams\\			
		\midrule
		\rowcolor{Gray}
		C1 & I1 & Templated Greenfield & Public Administration & 60 & $\approx$30 & 5\\
		C2 & I2 & Migration & B2B  E-Commerce & 8 & 10 & 3\\
		\rowcolor{Gray}
		C3 & I3 & Greenfield & IoT & 18 & 28 & 2 (up to 5)\\
		C4 & \multirow{2}{*}{I4} & Migration & B2B E-Commerce & 34 & $\approx$10 & 2\\
		C5 & & Migration & B2C E-Commerce & 8 & $\approx$10 & 2\\
		\rowcolor{Gray}
		C6 & I5 & Templated Greenfield & Logistics & 15-20 & 75 & $\approx$10\\
		\bottomrule
	\end{tabular}
\end{table*}

As shown \cref{tab:interviewcases}, we distinguish the cases into greenfield (new development from scratch), templated greenfield (new development based on legacy system), and migration (transformation of a monolithic legacy system into an MSA-based system) (Column Type). We further categorize each devlopment process by the domain of the microservice application under development (Column Domain). The number of microservices present in the application at the time of the interview (Column \#Services), number of people (Column \#Ppl) and teams (Column \#Teams) involved vary depending on the case. Case three is a special case; although there are formally only two teams, sub-teams are formed depending on the customizations to be performed to the microservice application, so that at certain points in time up to five teams work simultaneously on the application. In all cases the interviewees stated to apply the Scrum framework~\cite{Schwaber2002} for internal team organization. By contrast, the collaboration across teams was in all cases not following a particular formal methodology or model (cf.~\cref{subsec:MicroserviceDevProcess}). In addition, all interviewees reported that they strive for a DevOps culture~\cite{Ebert2016} in their SMOs. A detailed description of the cases can be found in our previous work~\cite{SorgallaInterviews2020}.

\subsection{Analytical Procedure}
For the analysis of the dataset, we used the Constant Comparison method~\cite{Shull2008}. That is, we rescreened existing paraphrases and marked challenges and/or solutions that our interviewees told us about with corresponding codes for challenges, obstacles, and solutions. We then used the coded statements across all cases to combine similar statements to higher-level challenges.

\subsection{Study Results and Challenges}\label{subsec:StudyResults}
Our analysis of the dataset resulted in the discovery of several common challenges across all cases. Comparable to other empirical studies, e.g.,~\cite{Taibi2020} or~\cite{Haselboeck2018}, our participants reported about the high technical complexity and high training effort during a microservice development process compared to a monolithic approach. 
Other discovered challenges in line with existing literature, e.g.,~\cite{Fritzsch2019}, concern the slicing of the business domain into individual microservices and the most suitable granularity of a microservice (cf.~\cref{sec:msa}).

In the following, we elaborate on two challenge areas (CA) which we found to be of particular concern for SMOs adopting a DevOps culture in more detail. 


\subsubsection*{CA1: Developing, Communicating, and Stabilizing a Common Architectural Understanding}
\label{subsubsec:challenge1understanding}
Developing a common architectural understanding of the architecture components of an application, especially about the goals and communication relationships of these components, in the case of MSA the microservices, is essential for developing a software in an organization which follows the DevOps paradigm~\cite{Bass2015}. The interviewees also think that the development of a general understanding of architecture among those involved in development is an important prerequisite for granting teams autonomy and trust. 

For cases C2, C4 and C5 (cf.~\cref{tab:interviewcases}), which each comprise approx. ten people and two to three teams, the practices to achieve this understanding are Scrum Dailies \cite{Schwaber2002} and regular developer meetings about the current status of the architecture. 
However, in case of more involved people, achieving a common understanding is reported to be very challenging. For cases C1, C3, and C6, the system development initially started with fewer people, and as the software product became successful, more people and teams were added. Regarding this development and the common architectural understanding I1 states that ``From one agile team to multiple agile teams is a huge leap, you have to regularly adapt and question the organization. [...] you need a common understanding of the architecture and a shared vision of where we want to go [...], we are working on that every day and I don't think we'll ever be done.''

A strategy that we observed to create this common architectural understanding in C1, C3, and C6 is the creation of new meeting formats. However, a contradicting key aspect of the DevOps culture is to minimize coordination across teams as much as possible~\cite{Bass2015}. The arising problem is also experienced by our interviewees. The more people and teams involved in exchanging knowledge to develop an architectural understanding, the more time-consuming the exchange becomes. In the case of C6, this has led to the discontinuation of comprehensive knowledge exchanges due to the excessive time involved. They now only meet on the cross-team level to discuss technologies, e.g., a particular authentication framework or a new programming language. We interpret this development as a step towards the introduction of horizontal knowledge exchange formats such as Guilds in the Spotify Model~\cite{Smite2019}. As a result, C6 is currently challenged with building a common understanding of the architecture only through these technology-focused discussions. This is a problem area that is also evident in the data of other empirical studies. For example, Bogner et al.~\cite{Bogner2019} report on the creation of numerous development guidelines by a large development organization to enforce a common architectural understanding. However, the development of guidelines requires that architecture decisions, technology choices, and use cases are documented~\cite{Haselboeck2017}, a practice we encountered only at C4 and C5.

In terms of technical documentation, the teams in all six cases use Swagger to document the microservices' APIs. Other documentation, such as a wiki system or UML diagrams, either is not used or not kept up to date. In almost all cases, access to the API documentation is not regulated centrally, but is instead provided by the respective team through explicit requests, e.g., by e-mail. Only C3 has extensive and organization-wide technical documentation as it is described by I3: ``Swagger is a good tool, but of course this is not completely sufficient, which is why we have an area where the entire concept of the IT platform {[}...{]} is explained. We also have a few tutorials.''

Summarizing CA1, we suspect that due to a mostly volatile organization, where the number of developers and software features often grows as development progresses, as well as the reported hard transition from a single to multiple agile teams, SMOs are particularly affected by the challenge of implementing a common architectural understanding as part of a successful DevOps culture. Documenting architecture  decisions, deriving appropriate guidelines, and an accessible technical documentation are key factors for an efficient development process that become more relevant with more teams and developers involved~\cite{Martini2013} and is therefore often not considered by SMOs early in the development process.  

\subsubsection*{CA2: Complexity of Deployment Techniques and Tools}
\label{subsubsec:challenge2ops}
A recurrent challenge we identified is how to deal with the operation of microservice applications within the development process. While cross-functional teams following the DevOps paradigm are mentioned in the literature, e.g.,~\cite{Nadareishvili2016}, as being recommended for the implementation of microservices architectures, in each of the researched cases we found specialized units for operating microservices instead of operators included as a part of a microservice team. In C1, C2, and C6 we encountered entire teams solely dedicated to operational aspects. In all cases, the development process included a handover of developed services to those specialized units for operating the microservice application. Although most interviewees were aware that this contradicts the ownership principle of microservices (cf.~\cref{subsec:MSAGeneral}) and they all stated to try to establish a pure DevOps without specialized teams, the effort to learn the basics of the necessary operational aspects is perceived as high. In this regard I2 comments ``The complexity (\textit{note: of cloud-based deployment platforms}) is already very, very high, you know. I would say that each of these functions in such a platform is a technology in itself that you have to learn.'' In contrast to operations, the SMOs are successful in including other professions, such as UI/UX, as parts of their cross-functional teams. Our data indicates that this is due to two main reasons. First, the inherently high complexity of the operational technologies and the associated high hurdle of learning and integrating them into the microservice development process. Second, the transfer of this knowledge not only into special units but into the individual microservice teams in order to do justice to a DevOps approach.

Summarizing CA2, deployment and operation in the SMOs studied is not in the responsibility of the teams to which the respective microservices belong. This seems to be due to the complexity of operation technologies and the associated learning effort. This might particularly be an issue for SMOs due to the challenging environment, where there are few resources to substitute, e.g., for a colleague who needs to learn an operation technology.

\section{Case Study}
\label{sec:casestudy}
In this section, we present a case study that we will use in the following sections to illustrate and validate our model-driven workflow (cf.~\cref{sec:workflow}) to address the challenges in \cref{sec:qualitative-study}. We decided for the usage of a case study to show the applicability of our approach because non-disclosure agreements prevent us from presenting our approach in the context of the explored SMO cases (cf.~\cref{subsec:interviewcases}). Therefore, we selected an open source case study microservice architecture, which maps to the design and implementation of the explored SMO cases w.r.t. the scope of our approach. More precisely, the case study (i) employs Swagger for API documentation (cf.~\cref{subsubsec:challenge1understanding}), (ii) uses synchronous and asynchronous communication means, (iii) is mainly based on the Java programming language, and (iv) the number of software components matches the smaller SMOs in our qualitative analysis (cf.~\cref{subsec:interviewcases}).

The case study is based on a fictional insurance company called Lakeside Mutual \cite{Stocker2018}. The application serves to exemplify different API patterns and DDD for MSA. The application comprises several micro-frontends~\cite{Peltonen2021}, i.e., \emph{se\-mi-in\-de\-pen\-dent fron\-tends} that invoke backend functionality, and microservices centered around the insurance sector, e.g., customer administration, risk management, and customer self-administration functions. The application's source code as well as documentation is publicly available on GitHub\furl{https://github.com/Microservice-API-Patterns/LakesideMutual}.

\begin{figure*}
    \centering
    \includegraphics[width=0.9\linewidth]{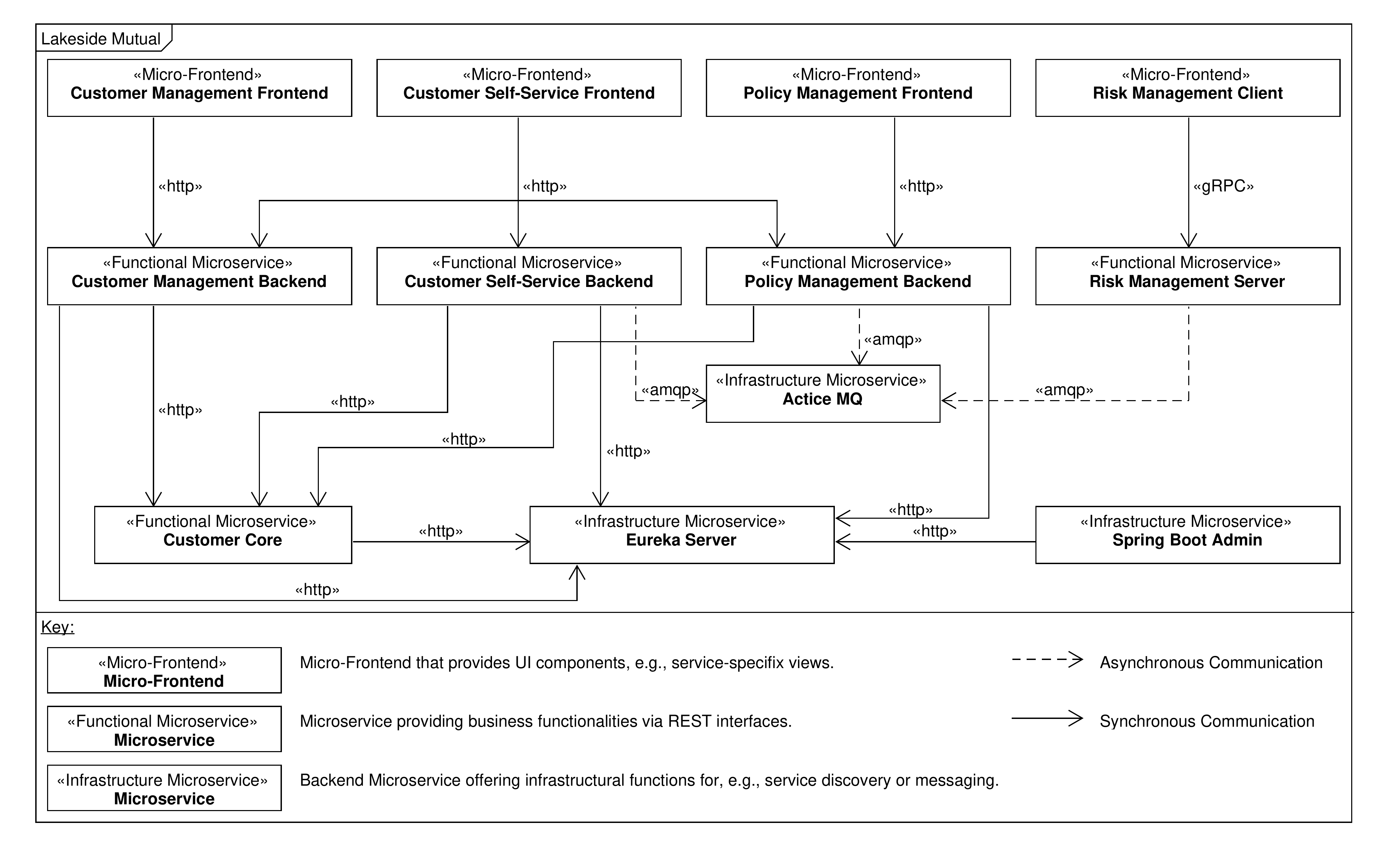}
    \caption{Structure of the case study Lakeside Mutual microservice application.}
    \label{fig:lakeside-mutual}
\end{figure*}

Figure~\ref{fig:lakeside-mutual} depicts the architectural design of the Lakeside Mutual application. Overall it consists of five functional backend microservice. Each microservice is aligned with a micro-frontend. 

Except for the \texttt{Risk Management Server}, all microservices are implemented in Java\furl{https://www.java.com} using the Spring framework\furl{https://spring.io}. A micro-frontend communicates with its aligned microservice using RESTful HTTP \cite{Fielding2000}. Additionally, the \texttt{Risk Management Client} and \texttt{Risk Management Server} communicate via \texttt{gRPC}. For internal service to service communication, the software system also relies on synchronous RESTful HTTP, but also on asynchronous \texttt{amqp} messaging over an \texttt{Active MQ} message broker. The \texttt{Customer Management Backend} and the \texttt{Customer Core} services also provide generated API documentations based on Swagger\furl{https://swagger.io/}.  

Besides the functional microservices, the Lakeside Mutual application also uses infrastructural microservices. The \texttt{Eureka} \texttt{Server} implements a Service Registry~\cite{Richards2016} to enable loose coupling between microservices and their different instances. For monitoring purposes, the \texttt{Spring} \texttt{Boot} \texttt{Admin} service provides a monitoring interface for the health status of individual services and the overall application. 

\section{A Model-Driven Workflow for Coping with DevOps-Related Challenges in Microservice Architecture Engineering}
\label{sec:workflow}
This section proposes a model-driven workflow based on LEMMA (cf.~\cref{sec:lemma}) to cope with the challenges identified in \cref{sec:qualitative-study}. More precisely, the workflow provides a common architectural understanding of a microservice application (cf. Challenge CA1 in \cref{subsubsec:challenge1understanding}), and reduces the complexity in deploying and operating microservice architectures (Challenge CA2).

In the following subsections, we present the design of the workflow (cf.~\cref{subsec:workflow}). Next, we describe the components, which we have added to LEMMA, to support the workflow. These components include (i) interoperability bridges between OpenAPI and LEMMA models (CA1; cf.~\cref{subsec:apispecifications}); (ii) an extension to the Service Modeling Language to allow the import of remote models (CA1; cf.~\cref{subsec:assemblingSystemModel}); (iii) a model processor to visualize microservice architectures (CA1; cf.~\ref{subsec:lemmaVisualizer}); (iv) enhancing the Operation Modeling Language (cf.~\cref{subsec:InfrastructureModels}; and (v) code generators for microservice deployment and operation (CA2; cf.~\cref{subsec:InfrastructureCodeGeneration}).

Furthermore, we present in detail prototypical components that we have added to the LEMMA ecosystem to support the workflow. These include deriving models from API documentation (cf.~\cref{subsec:apispecifications}) and assembling microservice models into an architecture model (cf.~\cref{subsec:assemblingSystemModel}) as a means to build a common architectural understanding (CA1), and enriching microservice models with deployment infrastructure models (cf.~\cref{subsec:InfrastructureModels}) as a means to more easily handle operational aspects for SMOs (CA2).

To ensure replicability of our results we have provided a GitHub repository\furl{https://github.com/SeelabFhdo/SN2021} which contains a documentation how to setup LEMMA and our in this article contributed extensions to it. It further contains all generated artifacts as well as sources and scripts to rerun the generations. Finally, it contains a manually created set of LEMMA models which represent all Java-based microservices of the Lakeside Mutual case study (cf.~\cref{sec:casestudy}).

\subsection{LEMMA-Based Workflow for Coping with DevOps Challenges}
\label{subsec:workflow}

\begin{figure*}[ht]
    \centering
    \includegraphics[width=0.7\linewidth]{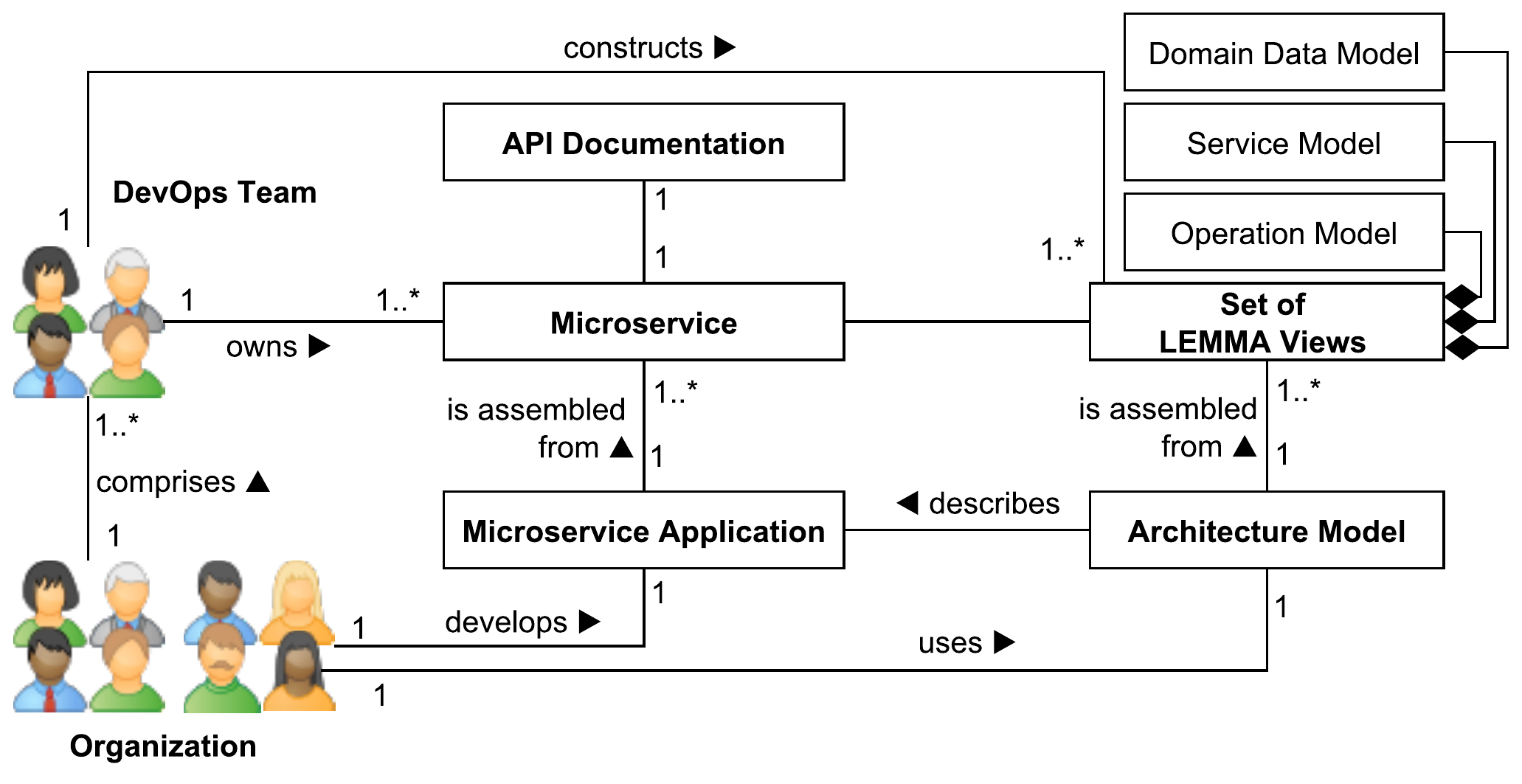}
    \caption{Overview of the concepts within the workflow and their interrelationships represented as a UML class diagram~\cite{OMG_UML_2017}.}
    \label{fig:workflow_structure}
\end{figure*}

\cref{fig:workflow_structure} shows the conceptual elements and their relationships which underlie the design of our LEMMA-based workflow for coping with the DevOps challenge areas (cf.~\cref{sec:qualitative-study}).

An \texttt{Organization} includes multiple \texttt{DevOps Teams}, each responsible for one or more \texttt{Microservices} (cf.~\cref{subsec:MicroserviceDevProcess}). The sum of all microservices forms the \texttt{Microservice Application} that is developed by the organization. Associated with a microservice is a corresponding documentation of its interfaces (\texttt{API Doc\-u\-men\-ta\-tion}). For each microservice owned by it, the team constructs a \texttt{Set of LEMMA Views} as a model representation (cf.~\cref{sec:lemma}). The sum of all LEMMA models forms an \texttt{Architecture Model} which describes the system's architecture. This model can be used by the organization, e.g., to gain insight into existing dependencies between the microservices involved.

\begin{figure*}[ht]
    \centering
    \includegraphics[width=\linewidth]{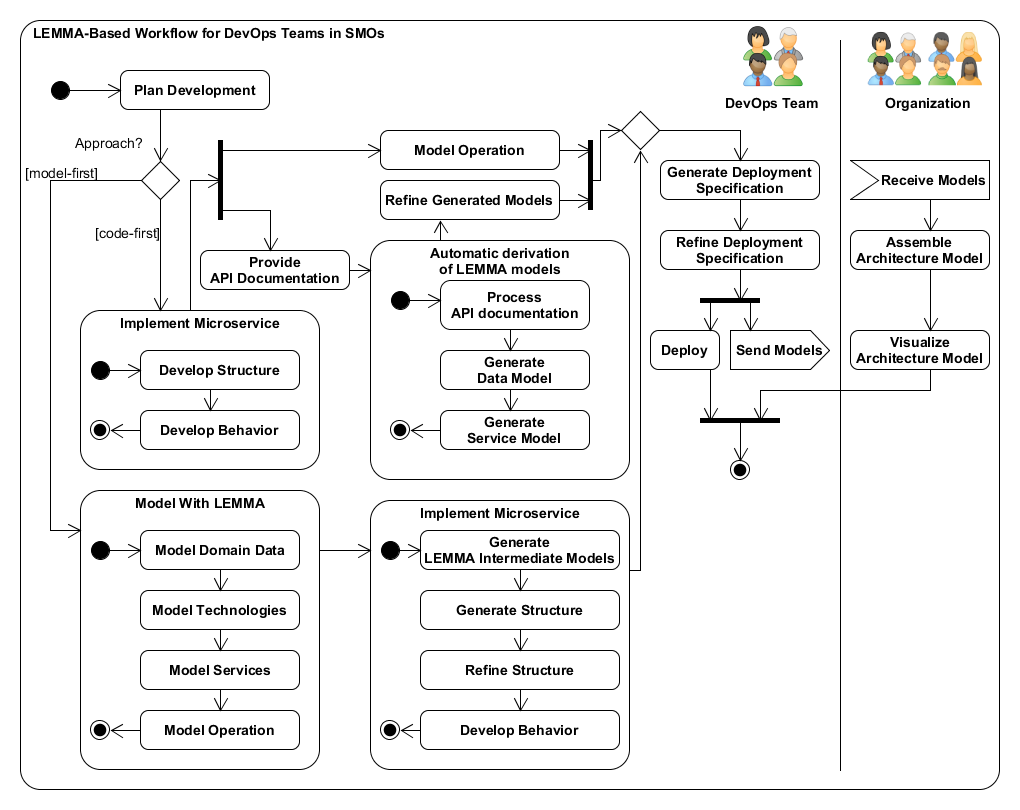}
    \caption{Proposed workflow for DevOps teams for model-driven microservices development represented as a UML activity diagram~\cite{OMG_UML_2017}.}
    \label{fig:workflow_flow}
\end{figure*}

Based on the conceptual elements and their relationships, \cref{fig:workflow_flow} shows our model-driven workflow for DevOps-based microservices development in SMOs as a UML acitivity diagram~\cite{OMG_UML_2017}.

We depict the workflow from the perspective of a single DevOps team including all steps required for the development of a new microservice. When incremental changes are made to individual aspects of a microservice, only the steps affected by the changes need to be performed.

The process starts with the planning of the development. The team decides whether to follow a \texttt{code-first} or \texttt{model-first} approach. We support both variants to allow the teams autonomy according to the DevOps paradigm~\cite{Bass2015}.
\paragraph{Code-First Approach}
Here, the team first implements the microservice consisting of \texttt{structure} and \texttt{behavior}. Based on the finished implementation, the team creates an \texttt{API Documentation}, which can done manually or automatically with tools such as Swagger\furl{https://swagger.io/}. Using the API documentation, a LEMMA domain model and a LEMMA service model are automatically derived (cf.~\cref{subsec:apispecifications}) and, if necessary, refined by the team. In parallel, the team creates a LEMMA operation model, since the information required for this kind of model cannot be derived from the API documentation (cf.~\cref{subsec:apispecifications}).
\paragraph{Model-First Approach} Alternatively, the team can decide to first model the structure and operation of the microservice using LEMMA. In the subsequent implementation activity, the structural aspects can be generated based on the previously constructed models and only the manual implementation of the behavior is necessary (cf.~\cref{sec:lemma}).

Regardless of which of the two approaches was chosen, at the end LEMMA domain, service, and operation models are available that describe the \emph{Dev} and \emph{Ops} aspects of the microservice under development. 

The operation model is then used to \texttt{Generate a Deployment Specification} for a container-based environment which mitigates the complexity of the operation (cf.~\cref{subsec:InfrastructureCodeGeneration}). The team refines this specification as needed and then deploys the microservice. In parallel, the models generated during the workflow are sent to a central model repository and made available to the entire organization where they can be used by other teams to gain insight and a common understanding of the application's architecture, e.g., by visualizing its structure.

Based on the use of model transformations and code generation steps, in the code-first as well as the model-first approach, we argue that the application of the workflow is possible with almost the same resources as the current development processes in the individual DevOps teams that we were able to explore as cases in the empirical study (cf.~\cref{sec:qualitative-study}). At its core, the code-first approach relies on the same development steps, i.e., implementing structure and behavior of a microservice, as non-model-based processes in the individual teams, so that even teams without experience in MDE can adapt the flow in a non-invasive way. Besides the actual implementation, the workflow provides a service's description in the form of LEMMA viewpoint models, which can be used as a communication basis and for knowledge transfer in order to create a common architecture understanding (CA1; cf.~\cref{subsubsec:challenge1understanding}) in the organization. This can be used to, e.g., accelerate verbal coordination processes between teams, improve the documentation, or identify microservice bad smells (\cite{Taibi2018}). In addition, by using LEMMA operation models and generating deployment specifications, it is easier for teams of an SMO to address the ops aspects themselves without passing on the responsibility for deployment to another unit (CA2; cf.~\cref{subsubsec:challenge2ops}). This enables teams to foster the ownership principle of MSA (cf.~\cref{subsec:MSAGeneral}).

\subsection{Derivation of Microservice Models from API Documentations}
\label{subsec:apispecifications}
To enable the model-driven workflow with sophisticated modeling support by LEMMA, we extended the ecosystem with the ability to derive data and service models from API documentation, that conforms to the OpenAPI Specification\footnote{Version 3.0.3} (OAS)~\cite{OpenAPI2020}, into LEMMA modeling files. OAS defines a standardized interface to describe RESTful APIs. One of the most popular tools implementing OAS is Swagger, which was used by all SMOs in the qualitative study (cf.~\cref{sec:qualitative-study}). 

\begin{figure*}
    \centering
    \includegraphics[width=0.8\linewidth]{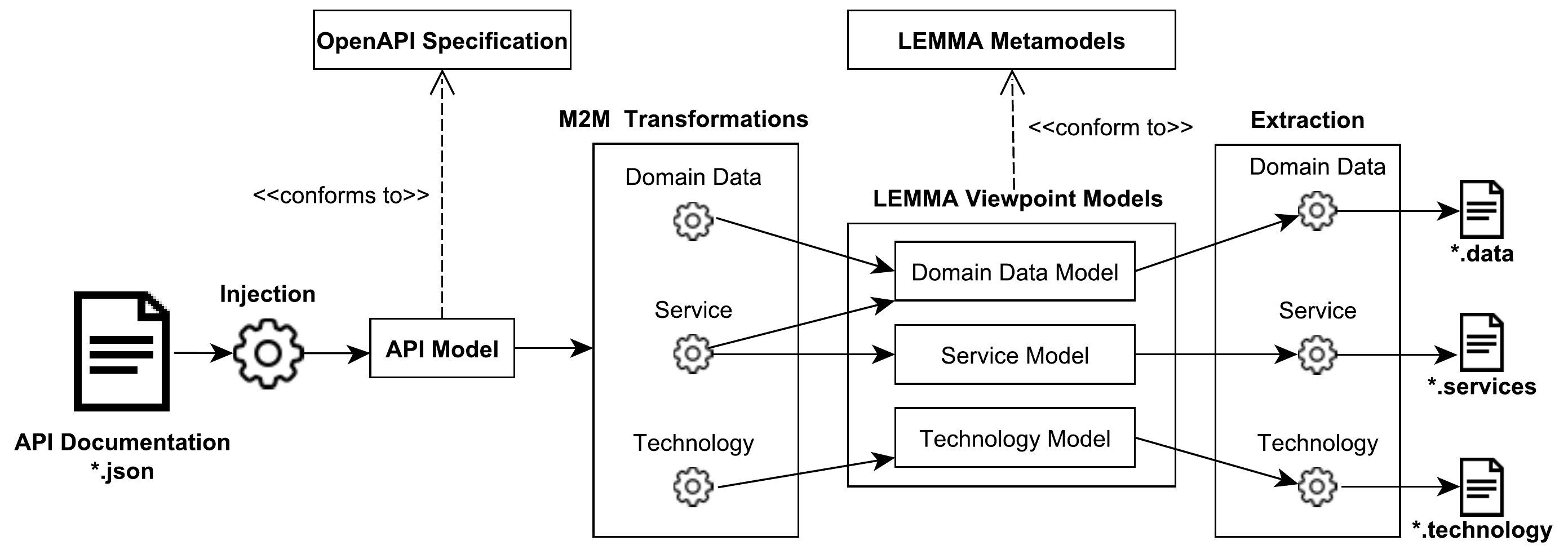}
    \caption{Process to transform OAS conform API models into LEMMA models.}
    \label{fig:openapi-interopability-bridge}
\end{figure*}

The transformation of OAS files into LEMMA files can be classified as an interoperability issue in which OAS models are to be converted into LEMMA models. We therefore applied the \emph{interoperability bridge} process proposed by Brambilla et al.~\cite{Brambilla2012}. \cref{fig:openapi-interopability-bridge} shows the applied interoperability bridge process.

First, using the Swagger parsing framework\furl{https://github.com/swagger-api/swagger-parser}, an OAS conform API model in the YAML~\cite{YAML_2009} or JSON~\cite{ECMA_JSON_2017} format is converted into an in-memory \texttt{API Model}. We then perform three model transformations where the information from the API Model is transformed into a \texttt{Do\-main Da\-ta Mod\-el}, \texttt{Ser\-vice Mod\-el}, and a \texttt{Tech\-nol\-o\-gy Mod\-el} which each correspond to their respective LEMMA metamodel (cf.~\cref{sec:lemma}). \cref{tab:openapi2lemma} describes which OAS objects\furl{https://spec.openapis.org/oas/v3.0.3\#openapi-object} are mapped to which LEMMA model kind.

\begin{table*}[htp]
\centering
\caption{Applied mappings between OAS and LEMMA concepts.}
\label{tab:openapi2lemma}
\begin{tabular}{L{0.15\textwidth} L{0.15\textwidth} L{0.65\textwidth}}
\toprule
\textbf{OAS Objects} & \textbf{LEMMA Model} & \textbf{Description of Mapping}\\
\midrule
\rowcolor{Gray}
Info & Domain Data & Name and version of the \texttt{Domain} and its \texttt{Context}\\

Schemas & Domain Data & \texttt{DataStructures} of the \texttt{Context}\\

\rowcolor{Gray}
Paths & Domain Data & Inline schemas are transformed to additional  \texttt{DataStructures} and arrays are mapped to \texttt{ListTypes}\\

\midrule

Info & Service & Name and commentary of the \texttt{Microservice}\\

\rowcolor{Gray}
Tag & Service & Used to derive a service's \texttt{Interfaces} which cluster \texttt{Operations}\\

Paths & Service & \texttt{Operations} of an \texttt{Interface} comprising its \texttt{Endpoints}, \texttt{Parameters}, HTTP request method as \texttt{Aspect}, and \texttt{Commentary}\\

\midrule

\rowcolor{Gray}
Media Types & Technology & All mentioned media types in a OAS model are transformed to \texttt{DataFormats} of the \texttt{RESTful HTTP Protocol}.\\
Data Types & Technology & Mapped to \texttt{TechnologySpecificPrimitiveTypes}.\\
\bottomrule
\end{tabular}
\end{table*}

To be able to transform the in-memory LEMMA models as files, we extended LEMMA with  \emph{extractors}~\cite{Brambilla2012} for technology, service, and data models.

\cref{lst:CustomerCoreOAS} and \cref{lst:CustomerCoreServiceModel} illustrate the application of the process. 

\begin{lstlisting}[language=JSON, caption=Excerpt of the OAS conform~ \texttt{Open\-API\-Cus\-tomer\-Core\-Back\-end\-Ser\-vice.json} describing the \texttt{Cus\-tomer\-Core} microservice., label=lst:CustomerCoreOAS, float]
"paths":{
  "/cities/{postalCode}":{
    "get":{
      "tags":["city-reference-data-holder"],
      "summary":"Get the cities for a postal code.",
      "operationId":"getCitiesForPostalCodeUsingGET",
      "produces":["*/*"],
      "parameters":[{
        "name":"postalCode",
        "in":"path",
        "description":"the postal code",
        "required":true,
        "type":"string"
      }],
      "responses":{
        "200":{
          "description":"OK",
            "schema":{"$ref":"#/definitions/CitiesResponseDto"}
        },
        /*...*/
\end{lstlisting}

\cref{lst:CustomerCoreOAS} shows an excerpt of the API documentation file of the Customer Core microservice from the case study (cf.~\cref{sec:casestudy}).
In detail, the listing presents the OAS description for a HTTP GET request on the path \texttt{cities/\{postCode\}} (Lines 2 and 3). This includes, e.g., the unique id \texttt{getCitiesForPostalCodeUsingGET} (Line 6) of the operation, the incoming parameters (Line 8 to 14), and the information that a response returns an object based on the \texttt{CitiesResponseDto} schema (Lines 15 to 19). The excerpt shows only the response for HTTP status code 200 (Line 16). OAS also offers the possibility to define responses for other status codes, e.g. HTTP status code 404, but these are currently not considered in the transformation to LEMMA in our prototypical implementation.

\begin{lstlisting}[caption=Excerpt of \texttt{cus\-tomer\-Core.ser\-vices} resulting from transforming the \texttt{Cus\-tomer\-Core} API documentation., style=lemma,  label=lst:CustomerCoreServiceModel, numbers=right, float]
import datatypes from "customerCore.data" 
  as CustomerCoreAPIData
import technology from "OpenApi.technology" as OpenApi

%%@technology%%(OpenApi)
public functional microservice 
  com.lakesidemutual.customercore.CustomerCore {
  interface cityReferenceDataHolder {
    ##---
    **Type** GET Operation for path /cities/{postalCode}
    **Summary** Get the cities for a postal code.
    **Description**##
    %%@required%% returnValue ##[INSERT PARAMETER DESC HERE]
    ---##
    %%@endpoints%%(OpenApi::_protocols.rest:
      "/cities/{postalCode}";)
    %%@OpenApi::_aspects.GetMapping%%
    getCitiesForPostalCodeUsingGET(
      sync in postalcode : string,  
      sync out returnValue : CustomerCoreAPIData::v100.
        CustomerCoreAPI.CitiesResponseDto);
  } 
  %%...%%
}
\end{lstlisting}

\cref{lst:CustomerCoreServiceModel} shows the LEMMA service model automatically transformed from the Cus\-tomer\-Core OAS model in \cref{lst:CustomerCoreOAS}. First, the results of the other transformations are imported into the service model. This includes the previously transformed LEMMA domain data model \texttt{cus\-tomer\-Core.data} resulting from the OAS schemas (Lines~1 and 2), which contains all data structures such as \texttt{Cities\-Response\-Dto}, and the technology model \texttt{OpenApi.tech\-nol\-o\-gy} (Line~3), which contains, e.g., the OpenAPI-specific primitive data types and the media types used in the CustomerCore OAS model. Line 4 enables the OpenApi technology for the \texttt{com.\-lake\-side\-mutual.\-cus\-tomer\-core.\-Cus\-tomer\-Core} microservice whose definition starts in Lines 6 and 7. The microservice comprises an interface named \texttt{city\-Reference\-Data\-Holder} which was derived by the associated tags in the OAS model (Line~8). The interfaces consists of the operation \texttt{get\-Cities\-For\-Postal\-Code\-Using\-GET} named after the OAS \texttt{operationId} (Lines 18 to 22). The operation commentary section (Lines 9 to 14) is populated using the \texttt{summary} information from OAS. The OAS \texttt{path} is added as an \texttt{endpoint} (Line~15) and the operation classified as an HTTP GET request (Line~17). The OAS response associated with the HTTP status code 200 is modeled as an OUT parameter and named \texttt{return\-Value} (Lines 20 and 21).


\subsection{Assembling a Common Architecture Model from Distributed Microservice Models}
\label{subsec:assemblingSystemModel}

Microservices need to interact with each other to realize coarse-grained functionality~\cite{Newman2015}. Thus, can depend on other services. In the case study (cf.~\cref{sec:casestudy}), such a relationship is found between the microservices \texttt{Customer Management Backend} and \texttt{Customer Core}. Such dependencies cannot be derived from an API documentation, since its purpose is to describe the provided interface of a service and not the invocation of functionality provided by other architecture components. However, these dependencies are essential in order to be able to assemble and assess an architecture model and to raise a common architectural understanding across the whole organization (cf.~\cref{subsubsec:challenge1understanding}). Therefore, within the workflow (cf.~\cref{subsec:workflow}), the dependencies should be added manually by the teams in the LEMMA models. This can be done during the \texttt{Model Services} activity when using the model-first approach and during the \texttt{Refine Generated Models} activity when using the code-first approach.

However, LEMMA service models originally were only able to depend on other LEMMA service models if they are accessible in the local file system. Therefore, to allow teams the expression of interaction dependencies with the microservices of other teams, we have extended LEMMA to allow external service imports. \cref{lst:CustomerManagementServiceModel} shows the service model of the \texttt{Customer Management Backend} microservice from the case study. The microservice imports show the two alternatives. The syntax for importing locally accessible service files is shown in Lines 3 to 4. Alternatively, the import in Lines 6 to 8 exemplifies the mechanism for external imports.

\begin{lstlisting}[caption=Excerpt of  \texttt{customerManagementBackend.services} illustrating the external import feature., style=lemma,  label=lst:CustomerManagementServiceModel, numbers=left, float]
import datatypes from "customerManagementBackend.data" 
  as customerManagementBackend
import microservices from "../customer-core/customerCore
  .services"  as customerCoreServices
//External import as alternative to Lines 3 to 4
import microservices from "https://repo.lakeside.com/
  teamB/customercore.json" to "../customer-core/
  customerCore.services" as customerCoreServices

public functional microservice
  com.lakesidemutual.customerManagementBackend
  .CustomerManagementBackend {
    required microservices {customerCoreServices::com.
    lakesidemutual.customercore.CustomerCore}
    %%...%%
\end{lstlisting}

As soon as the Eclipse IDE detects such an external import in the model, it offers a quickfix that automatically downloads the referenced file and, if it is OAS-compliant API documentation, starts a corresponding transformation to LEMMA (see Section 3.2). This also makes it possible to model a dependency to a service of another team, even if this team does not yet provide its own model but only API documentation. 

Since LEMMA models are textual~\cite{Rademacher2020} and with the extension it is possible to import external sources, the model files of the different teams can be managed centrally as an architecture model by a version management system such as Git and thus integrated into CI/CD pipelines, e.g., by a Git hook\footnote{https://git-scm.com/book/en/v2/Customizing-Git-Git-Hooks} that copies the models to a central model repository with each release of the microservice.

\subsection{Visualization of Microservice Architecture Models}
\label{subsec:lemmaVisualizer}
\begin{figure*}
    \centering
    \includegraphics[width=0.8\linewidth]{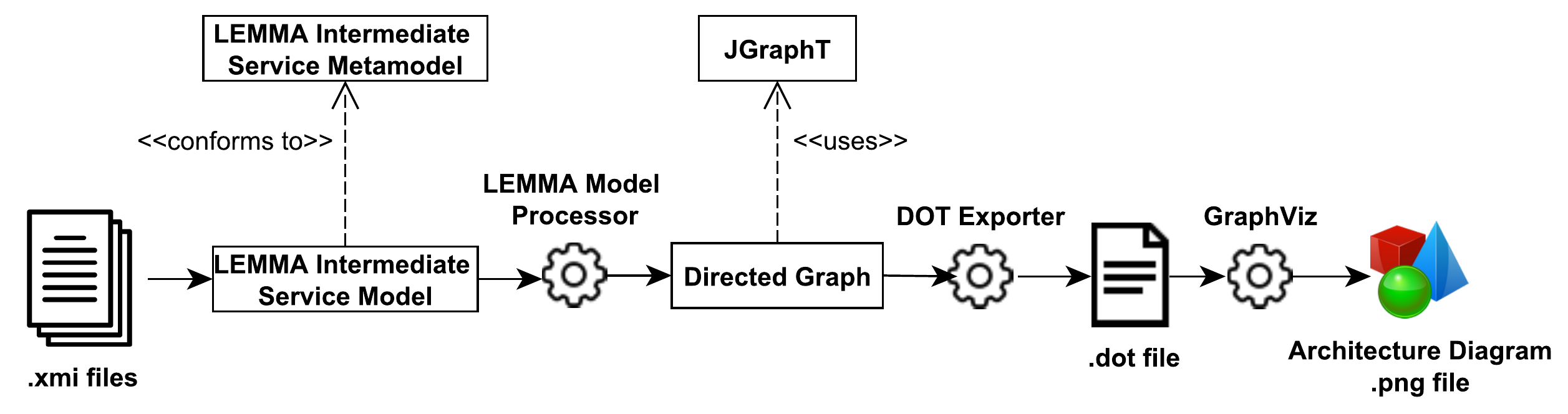}
    \caption{Model-to-Text transformation to generate a visual representation of a microservice application's architecture.}
    \label{fig:lemmaVisualizerTransformation}
\end{figure*}

To enable visualization of the architecture using LEMMA (cf.~\cref{subsec:workflow}), we have developed the LEMMA Visualizer\furl{https://github.com/SeelabFhdo/SN2021/blob/master/de.fhdo.lemma.visualizer-0.8.0-SNAPSHOT-standalone.jar}. It is able to transform several LEMMA intermediate service models (cf.~\cref{subsubsec:intermediatemodelprocessing}) into a single graphical representation using a \emph{model-to-text transformation}~\cite{Combemale2017}. The steps of the transformation are depicted in \cref{fig:lemmaVisualizerTransformation}.

\begin{figure*}
    \centering
    \includegraphics[width=\linewidth]{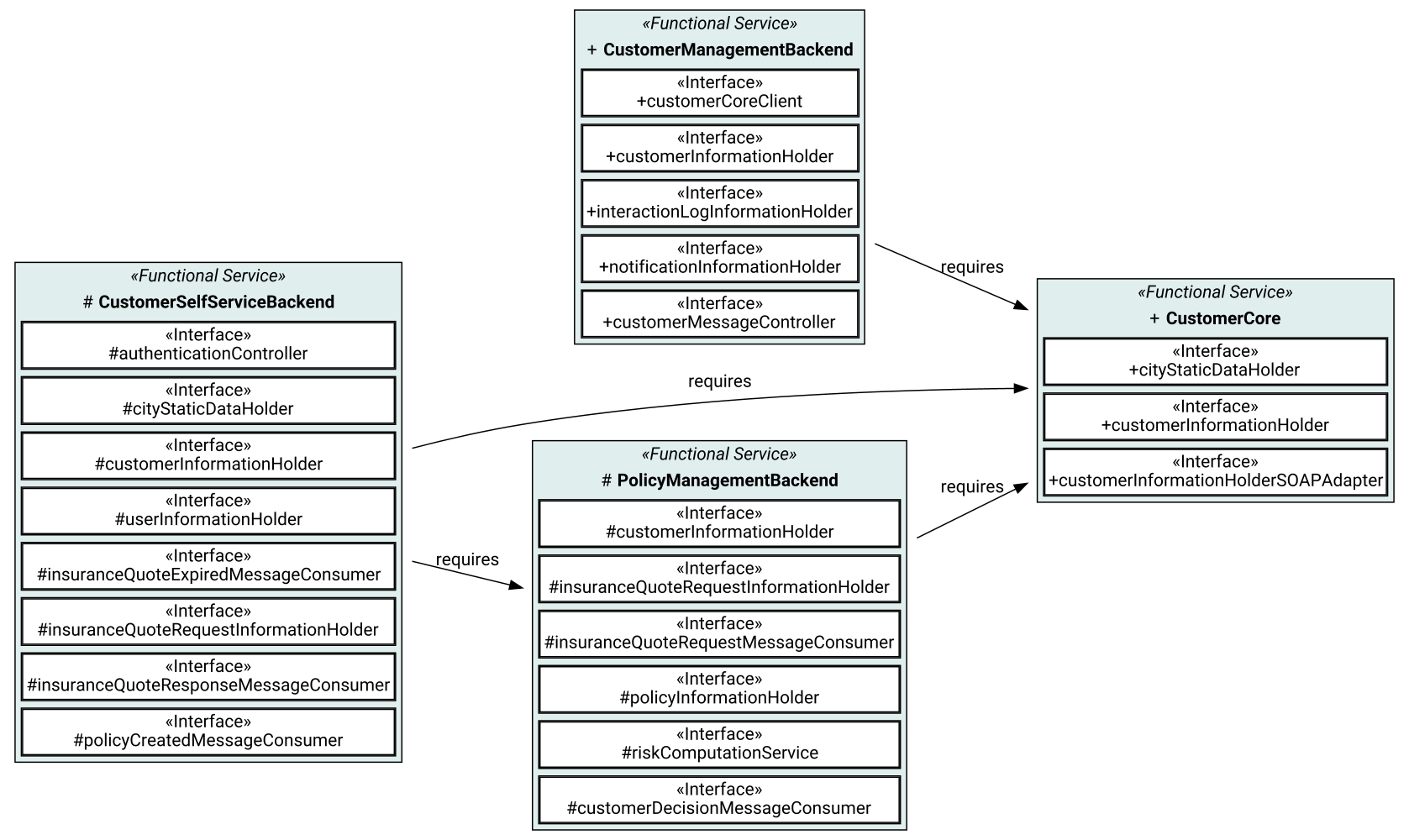}
    \caption{Generated visualization of the architecture of the case study (cf.~\cref{sec:casestudy}).}
    \label{fig:lemmaVisualizerResult}
\end{figure*}

First, one or more intermediate service model files are passed to the LEMMA Model Processor (cf.~\cref{subsubsec:intermediatemodelprocessing}). These are converted to their in-memory representation and then processed. Using the JGraphT framework~\cite{Michail2020}, we create a directed graph that we populate with microservices found during the processing of the intermediate models as vertices and any existing imports from other services as edges. Microservices that are neither imported nor import another service and are thus without an edge are added as isolated vertices. Then we use JGraphT's DOTExporter to convert the graph into a textual representation of the graph based on the DOT language\footnote{\url{https://graphviz.org/doc/info/lang.html}}. During the export, we enrich the DOT representation with attributes which describe the appearance of the vertices and edges for the later visualization, e.g., we add coloring and describe vertex shapes.  Finally, we use GraphViz~\cite{Ellson2002} to generate an image of the graph's DOT representation that represents the system architecture in the form of a box-and-line diagram as a Portable Network Graphics (\texttt{.png}) file. The visualizer supports the setting of different levels of detail of the display in relation to the attributes of microservice vertices. A resulting architecture image which shows the functional microservices from the case study with a detail level which shows interfaces but not operations is shown in \cref{fig:lemmaVisualizerResult}.

\subsection{Enhancing Distributed Microservice Models with Deployment Infrastructure Models}\label{subsec:InfrastructureModels}

In this subsection, we elaborate on the creation of an operation model (cf.~\cref{sec:lemma}) that complements the previously created data and service models to form a complete set of LEMMA views describing a microservice (cf.~\cref{subsec:workflow}), by using LEMMAs Operation Modeling Language (OML) \cref{subsec:MicroserviceArchitectureModelConstruction}). This approach specifically addresses the Operation and Deployment Stages of MSA (cf.~\cref{subsec:OperationStage} and \cref{subsec:DeploymentStage}) and therefore addresses CA2 (cf.~\cref{subsec:StudyResults}) by providing functionalities for describing the microservice models' deployment, including their dependencies to infrastructural services, e.g., API gateways, services discoveries, and databases. Additionally, OML abstracts from concrete technology-specific deployment configurations and reduces the overall complexity of deploying a microservice application. 

In order to enable DevOps teams in SMOs to take full ownership of their respective services, which mitigates the need to apply specialized teams dedicated to operating the whole microservice application (cf.~\cref{sec:qualitative-study}), we have extended the OML with means to import other operation models as \emph{nodes} and, therefore, nest operation specifications with each other. I.e. teams do not have to maintain individual models for infrastructure microservices, but can use the new mechanism to import the operation model, e.g., for a Eureka service discovery (cf.~\cref{sec:casestudy}), from a central model repository (cf.~\cref{subsec:assemblingSystemModel}).

OML now enables the DevOps Team to describe the deployment of a microservice and all necessary dependencies. \cref{lst:CustomerCoreOperationModel} shows an excerpt of the operation model for the deployment of the \texttt{Cus\-tomer\-Core} microservice. Lines 1 and 2 of the listing imports the \texttt{cus\-tomer\-Core.services} model derived from the services' Open API specification (cf.~\cref{subsec:apispecifications}). The following Lines 3 to 6 are dealing with the import of the technology for service deployment. The \texttt{Container\-\_base} technology model uses Docker\furl{https://www.docker.com/} and Kubernetes\furl{https://kubernetes.io/} for service deployment. Lines 7 and 8 illustrates the new possibility to import other operation models as nodes by importing the \texttt{eureka.operation} model that describes the deployment of a service discovery by the Eureka\furl{https://github.com/Netflix/eureka} technology.

\begin{lstlisting}[caption=Excerpt of the \texttt{customerCore.operation} model from our case study., style=lemma, label=lst:CustomerCoreOperationModel, numbers=left, float]
import microservices from "customerCore.services" 
  as customerCoreServices
import technology from "../technology/
  container_base.technology" as container_base
import technology from "../technology/
  javaWithSpring.technology" as protocolTechnology
import nodes from "../eureka-server/
  eurekaServer.operation" as eureka

%%@technology%%(container_base)
%%@technology%%(protocolTechnology)
container CustomerCoreContainer deployment technology 
  container_base::_deployment.Kubernetes
  deploys customerCoreServices::com.lakesidemutual.
  customercore.CustomerCore depends on nodes 
  eureka::ServiceDiscovery {
      customerCoreServices::com.lakesidemutual
      .customercore.CustomerCore {
        eurekaUri = "http://localhost:8761" 
          basic endpoints {
        protocolTechnology::_protocols.rest: 
          "http://localhost:8110"; }
    }
  }
\end{lstlisting}

Lines 10 and 11 assign the technology to the \texttt{Cus\-tomerCoreCon\-tain\-er} (Line 12). The \texttt{container} acts as a vessel for the deployed microservices and clusters deployment-relevant information, e.g., dependencies to infrastructural components such as databases, service-specific configurations, and protocol-specific endpoints. For this purpose, Line 12 and 13 create the \texttt{Cus\-tomerCoreCon\-tain\-er} and assign the \texttt{Kubernetes} deployment technology which is imported from the \texttt{container\-\_base} technology model. The deployment of the \texttt{Cus\-tomer\-Core} microservice into the container is shown in Lines 14 and 15. The following Lines 15 to 24, show the dependency to the \texttt{Ser\-vice\-Dis\-cov\-ery} imported from the \texttt{eu\-reka.op\-er\-a\-tion} model. In detail, Line 19 includes the service-specific configuration of the \texttt{Cus\-tomer\-Core} microservice by specifying the \texttt{eu\-rekaUri} responsible for configuring the dependency to the \texttt{Service\-Discovery}. The \texttt{Cus\-tomer\-Core} microservice exposes its functionality via a \texttt{rest} endpoint as defined in Lines 20 to 22.

Besides modeling the deployment of microservice-specific configurations, OML also enables the DevOps team to specify infrastructural components' deployment, e.g., service discoveries and databases. \cref{lst:ServiceDiscoveryOperationModel} describes the deployment of the \texttt{Ser\-viceDis\-cov\-ery}. Lines 1 to 3 are responsible for importing the \texttt{cont\-ain\-er\-base.\-tech\-nol\-o\-gy} and \texttt{eureka.tech\-nol\-o\-gy} models. The models include the specification of the technology used for the deployment of the \texttt{Ser\-vice\-Dis\-cov\-ery}. Lines 4 and 5 import the \texttt{Cus\-tomer\-Core} service which uses the service discovery. Lines 7 and 8 assign the imported technology to the \texttt{Ser\-vice\-Dis\-cov\-ery}.

\begin{lstlisting}[caption=Excerpt of the service discovery operation model from our case study., style=lemma, label=lst:ServiceDiscoveryOperationModel, float]
import technology from "docker.technology" 
  as containerTechnology
import technology from "eureka.technology" as Eureka
import nodes from "customerCore.operation" 
  as customerCore

%%@technology%%(containerTechnology)
%%@technology%%(Eureka)
ServiceDiscovery is Eureka::_infrastructure.Eureka
  used by nodes customerCore::CustomerCoreContainer {
  %%...%%
  default values {
    hostname="DiscoveryService"
    port = 8761
    %%...%%
  }
}
\end{lstlisting}

Line 9 starts the actual specification of the \texttt{ServiceDiscovery}, which uses the imported \texttt{Eureka} technology. The following Line 10 contains the dependency to the \texttt{CustomerCoreContainer}, specified in  \cref{lst:CustomerCoreOperationModel}. The service-specific configuration of the \texttt{ServiceDiscovery} is set via the assignment of \texttt{default values} in Lines 12 to 16. Line 13 and 14 set the actual \texttt{hostname} and \texttt{port} of the service. 

Overall, LEMMA's OML enables the DevOps team to construct operation models which specify the deployment of microservices and their dependencies on the microservice application's infrastructural components. The operation models consist of the concepts of containers and infrastructure nodes. Containers (cf.~\cref{lst:CustomerCoreOperationModel}) specify the deployment of microservice, whereby infrastructure nodes contain the configuration for infrastructural components, e.g., API gateways, databases, and service discoveries (cf.~\cref{lst:ServiceDiscoveryOperationModel}). 

\subsection{Generating Code from Distributed Deployment Infrastructure Models}
\label{subsec:InfrastructureCodeGeneration}

\begin{figure*}
    \centering
    \includegraphics[width=\linewidth]{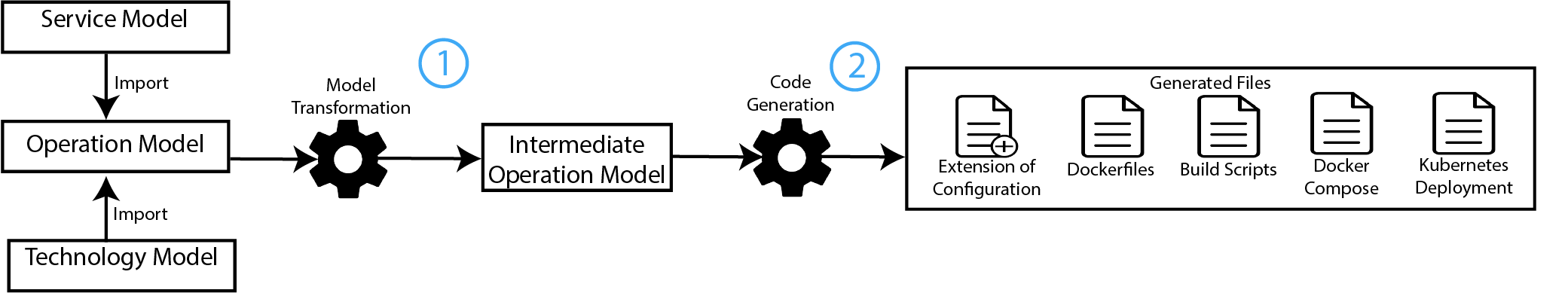}
    \caption{Code generation pipeline for creating deployment relevant artifacts from LEMMAs operation models.}
    \label{fig:code-generation-pipeline}
\end{figure*}

In \cref{subsec:InfrastructureModels} we introduced OML as a methodology to describe the deployment of a service-based software system. In this subsection, we contribute a code generation pipeline for creating deployment-related artifacts based on the operation models using LEMMA's Model Processor (cf.~\cref{subsubsec:intermediatemodelprocessing}). As depicted in \cref{fig:code-generation-pipeline}, the code generation pipeline consists of two consecutive stages. 

The first stage of the code generation pipeline consists of a \emph{model-to-model transformation}~\cite{Combemale2017} transforming an operation model into an intermediate operation model in the sense of LEMMA's intermediate model processing (cf.~\cref{subsubsec:intermediatemodelprocessing}).  

The second stage of the code generation pipeline deals with the creation of the deployment-relevant artifacts. Based on an intermediate operation model, the code generators already included in LEMMA (cf.~\cref{sec:lemma}) provide a variety of different functionalities that are usually bound to a specific technology model. As already shown in \cref{lst:CustomerCoreOperationModel} and \cref{lst:ServiceDiscoveryOperationModel}, the described operation models both use the \texttt{con\-tain\-er\-\_base} technology model.  

The \texttt{con\-tain\-er\-\_base} model clusters a technology stack suited for a service-based software system with focusing on container technologies~\cite{Kang2016} such as Docker, Docker-Compose, and Kubernetes. \cref{lst:ContainerBaseTechnologyModel} shows an excerpt of this specific technology model. Line 1 specifies the actual name of the model. Line 2 to 6 describe the \texttt{deployment technologies} of the model, in this particular case \texttt{Kubernetes}. Additionally, \texttt{Kubernetes} supports the \texttt{operation environments} \texttt{golang}, \texttt{python3}, and \texttt{openjdk} as its default.

\begin{lstlisting}[caption=Excerpt of the container\-\_base technology model (\texttt{docker.tech\-nol\-o\-gy}) for gen\-er\-at\-ing de\-ploy\-ment \-re\-lat\-ed ar\-ti\-facts., style=lemma, label=lst:ContainerBaseTechnologyModel,  numbers=left, float]
technology container_base {
  deployment technologies {
    Kubernetes {
      operation environments="openjdk:11-jdk-slim" 
      default, "golang", "python3";}
    %%...%%
  }
  operation aspects {
    //Aspect for configuring custom dockerfiles
    aspect Dockerfile<singleval> for containers, 
    infrastructure {string contents <mandatory>;}
    %%...%%
  }
}
\end{lstlisting}

The second part of \cref{lst:ContainerBaseTechnologyModel} contains the definition of operation aspects for further service deployment specification from Lines 8 to 13. Lines 10 and 11 define the \texttt{Dockerfile} aspect, which can be applied to \texttt{containers} in operation models. The aspect consists of a single attribute named \texttt{content} containing the actual content of the Dockerfile. Furthermore, the \texttt{content} attribute has the property \texttt{mandatory} to it, so it can only be configured a single time per container. 

The container\-base code generator\furl{https://github.com/SeelabFhdo/lemma/tree/master/code\%20generators/de.fhdo.lemma.model_processing.code_generation.container_base}, compatible with the eponymous technology model, creates a set of different deployment-related artifacts such as build scripts, Dockerfiles, Kubernetes files, and extends existing service configuration files. Based on the operation an technology models from \cref{lst:ContainerBaseTechnologyModel} and \cref{lst:CustomerCoreOperationModel} the code generation pipeline creates executable configuration from the start without any additional configuration needed.

\cref{lst:CustomerCoreDockerfile} shows a Dockerfile created by the container\-base generator from the described operation models. The Dockerfile contains a basic configuration consisting of a docker image deducted from the operation environment configured in the operation model. In Lines 2 to 15 several artifacts are copied into the image. Line 17 describes the port 8110 on which the microservice is started. Finally, Lines 18 to 20 define the \texttt{entrypoint} of the docker image to compile and run the microservice.

\begin{lstlisting}[caption=Dockerfile for the customer core service created by the code generation pipeline. , style=dockerfile, label=lst:CustomerCoreDockerfile, numbers=right, float]
### syntax=docker/dockerfile:experimental##
FROM maven:3.6.1-jdk-8-alpine AS build
ARG BASE=/usr/src/app
COPY pom.xml ${BASE}/
COPY src ${BASE}/src
RUN --mount=type=cache,target=/root/.m2 
  mvn -f ${BASE}/pom.xml install -DskipTests

FROM openjdk:8-jdk-alpine
COPY --from=build /usr/src/app/target
  /dependency/BOOT-INF/lib/* /app/lib/
COPY --from=build /usr/src/app/target
  /dependency/META-INF /app/META-INF
COPY --from=build /usr/src/app/target
  /dependency/BOOT-INF/classes /app

EXPOSE 8110
ENTRYPOINT ["java","-cp","app:app/lib/*",
  "com.lakesidemutual.customercore.
  CustomerCoreApplication"]
\end{lstlisting}

The container\-base generator creates a basic Dockerfile configuration to create a more advanced configuration for a Dockerfile. OML provides the mechanism of operation aspects to create custom a  Dockerfile. Furthermore, the operation aspect mechanism also applies for Docker-Compose and Kubernetes configurations. 

In Addition to the Dockerfiles, the container\-base generator also creates Kubernetes deployment files. Generally, the Kubernetes file consists of the deployment and service parts. The deployment part described in \cref{lst:KubernetesDeployment} contains the configuration of the Kubernetes pod\furl{https://kubernetes.io/docs/concepts/workloads/pods/} the microservice gets deployed to. Line 1 defines the \texttt{apiVersion} the Kubernetes file uses. The following Line 2  contains the definition of the configuration \texttt{kind: deployment}. Line 5 to 7 specifying a configuration type overarching name to the deployment, in this case, \texttt{customercorecontainer}. Line 8 indicates the configuration of the Kubernetes deployment, specifically the set of \texttt{replicas} that should be created for the deployment.

\begin{lstlisting}[caption=Excerpt of the deployment part of the Kubernetes file for the CustomerCore service. , style=kubernetes, label=lst:KubernetesDeployment, numbers=left, float]
apiVersion: apps/v1
kind: Deployment
metadata:
  ...
  labels:
    app: customercorecontainer
  name: customercorecontainer
spec:
  replicas: 1
  selector:
    matchLabels:
      app: customercorecontainer
...
\end{lstlisting}

\cref{lst:KubernetesService} contains the service part of the Kubernetes deployment and contains the configuration on how the microservices application is exposed. As previously, Lines 1 to 2 contain the information about the \texttt{apiVersion} and configuration type of the Kubernetes file \texttt{kind: Service}. Followed by the name assignment of the deployment in Lines 3 to 7. The listing defines the exposure of the microservice via port \texttt{8110} in Lines 9 to 12. 

\begin{lstlisting}[caption=Excerpt of the service part of the Kubernetes file for the CustomerCore service. , style=kubernetes, label=lst:KubernetesService,  numbers=left, float]
apiVersion: v1
kind: Service
metadata:
  labels:
    app: customercorecontainer
  name: customercorecontainer
spec:
  ports:
  - name: "8110"
    port: 8110
    targetPort: 8110
...    
\end{lstlisting}

Supplementary to the deployment-related generated artifacts, LEMMA's code generation pipeline also supports the extension of existing service configurations. For this purpose, we implemented additional code generators for technologies, e.g., MongoDB\furl{https://www.mongodb.com/}, MariaDB\furl{https://mariadb.org/}, Zuul, and Eureka. \cref{lst:CustomerCoreOperationModel} shows in Line 13 a property for specifying the \texttt{eurekaUri}. Based on this property, the spring \emph{spring eureka code generator} extends the service's configuration in the specified \texttt{CustomerCoreContaier}. 

\cref{lst:property} contains a variety of configurations for Spring-based microservice implementations. The \texttt{spring.\-application.\-name} and \texttt{server.\-port} in Lines 1 and 2 are derived from the modeled microservice's name and its specified endpoint from the LEMMA models. Lines 3 to 7 are deduced from the Eureka configuration shown in \cref{lst:CustomerCoreOperationModel}. They configure the endpoints for connecting to the eureka service discovery.

\begin{lstlisting}[caption=Excerpt of the \texttt{CustomerCoreService} configuration file. , style=property, label=lst:property, numbers=right, float]
spring.application.name=CustomerCoreService
server.port=8091
eureka.instance.preferIpAddress=true
eureka.client.registerWithEureka=true
eureka.client.fetchRegistry=true
eureka.client.serviceUrl.defaultZone=
  ${EUREKA_URI:http://discovery-service:8761/eureka}
\end{lstlisting}

\section{Validation}
\label{sec:validation}

In this section, we validate the present LEMMA extensions that implement the workflow  (cf.~\cref{sec:workflow}). To enable replicability of our results, we provide a validation package on GitHub\furl{https://github.com/SeelabFhdo/SN2021}. 

In order to make the validation feasible, we first reconstructed the functional backend and infrastructure microservices of Lakeside Mutual (cf.~\cref{sec:casestudy}) using a systematic process~\cite{Rademacher2020b}. This step was necessary because the backend and infrastructure microservices of Lakeside Mutual are implemented in Java and not modeled with LEMMA. In detail, our reconstruction includes all four Java-based functional microservices and the infrastructural microservices \texttt{Eureka Server} and \texttt{Spring Boot Admin} (cf.~\cref{sec:casestudy}).

In addition, we retrieved the current API documentation of Lakeside Mutual by putting the architecture into operation and triggering the generation of the documentation using prepared REST requests. At the end of this process, we could refer to the current API documentations of Lakeside Mutual's \texttt{Customer} \texttt{Core} and \texttt{Customer} \texttt{Management} \texttt{Backend} (cf.~\cref{fig:lakeside-mutual} in \cref{sec:casestudy}), which are the two components for which the application provides API documentation.

We then performed the individual generation steps of our workflow (cf.~\cref{sec:workflow}) based on our reconstructed LEMMA models and the case study's API documentation. We illustrate the results of the application of our workflow as shown in \cref{tab:LoCval} using the Lines of Code (LoC) metric.

\begin{table*}[htp]
\centering
\caption{Overview of the number of LoC of the different model artifacts involved in the LEMMA-based workflow.}
\label{tab:LoCval}

\begin{tabular}{L{0.4\textwidth} L{0.2\textwidth} L{0.2\textwidth} R{0.1\textwidth}}
\toprule
\textbf{Service} & \textbf{Type} & \textbf{Viewpoint} & \textbf{LoC}\\
\midrule
\textbf{Manually Built Models} & & & \\
\rowcolor{Gray}
All Services & LEMMA & All & \zvalLocManModelsLm{}\\
All Services & LEMMA & Operation & \zvalLocManModelsLmOperation{}\\
\rowcolor{Gray}
Customer Core & LEMMA & All & \zvalLocManModelsCustomerCore{}\\
Customer Core & LEMMA & Operation & \zvalLocManModelsCustomerCoreOperationLEMMA{}\\
\rowcolor{Gray}
Customer Management Backend & LEMMA & All & \zvalLocManModelsCustomerManagement{}\\
Customer Management Backend & LEMMA & Operation & \zvalLocManModelsCustomerManagementOperationLEMMA{}\\
\midrule
\textbf{API Documentation} & & & \\
\rowcolor{Gray}
Customer Core & OAS/JSON &  Domain Data \& Service  & \zvalLocOpenApiCustomerCoreJSON{}\\
Customer Core & LEMMA & Domain Data \& Service & \zvalLocGenModelsCustomerCore{}\\
\rowcolor{Gray}
Customer Management Backend & OAS/JSON &  Domain Data \& Service & \zvalLocOpenApiCustomerManagementJSON{}\\
Customer Management Backend & LEMMA & Domain Data \& Service &  \zvalLocGenModelsCustomerManagement{}\\
\midrule
\textbf{Deployment Specification} & & & \\
\rowcolor{Gray}
All Services & Docker\newline{}Kubernetes & Operation & \zvalLocDeploymentSpecs{}\\
\bottomrule
\end{tabular}
\end{table*}

As \cref{tab:LoCval} shows, using the OAS-conform API documentation, we were able to generate 171 and 174 LoC of LEMMA Domain Data and Service files for the Customer Core and the Customer Management Backend microservices, respectively. Although the same operations and parameters for interfaces are present in the models generated by our workflow and the reconstructed LEMMA models, the LoC are higher in our reconstructed models. This is due to the fact that, e.g., the operation-related portion of LoC or technology-related annotations for databases are present in the manual models, but not in the generated ones, since no information on this is available from the API documentation. 

Regarding the generation of deployment specifications, we were able to generate 285 lines of infrastructure code for Docker and Kubernetes from the reconstructed operation models of the functional microservices. Teams can abstract from technology-specific infrastructure code and, in combination with LEMMA's source code generators such as the Java Base Generator~\cite{Rademacher2020c}, generate directly executable and deployable stubs of their services. 

\section{Discussion}
\label{sec:discussion}
The model-based workflow presented in \cref{sec:workflow} addresses the previously identified challenge areas (cf.~\cref{sec:qualitative-study}). In detail, the workflow provides means to establish a common understanding of architecture in an organization scaling to the level of multiple teams for the first time (CA1) and the complexity of operational aspects in microservice engineering (CA2).  

We argue that by documenting the architecture in a centralized manner (cf.  \cref{subsec:apispecifications} and \cref{subsec:assemblingSystemModel}), combined with the ability to visualize it (cf.~\cref{subsec:lemmaVisualizer}), teams and higher-level stakeholders, such as project sponsors, have a good basis for sharing knowledge and gaining insight into each other's development artifacts through the inherent abstraction property of the models~\cite{Ludewig2003}. \emph{Box-and-line} diagrams, in particular, have the advantage that people can more easily grasp relations between concepts~\cite{Combemale2017}.

Another added value of our approach is the ability to seamlessly integrate deployment specifications into architecture models as a LEMMA operation model with the possibility to derive deployment configurations for heterogeneous deployment technologies, i.e., to generate them for Docker and Kubernetes (cf.~\cref{subsec:InfrastructureCodeGeneration}). To this regard, Combemale et al.~\cite{Combemale2017} underline the added value of models to abstract complexity in the deployment process making the process more manageable. However, deployment technologies supported by our workflow constitute de-facto standards~\cite{Shah2019}, LEMMA does limited justice to the heterogeneous technology landscape concerning cloud providers. In particular, we do not specifically address cloud-based deployment platforms such as AWS\furl{https://aws.amazon.com/} or Azure\furl{https://azure.microsoft.com/}. Presumably, LEMMA is able to support such technologies through specific technology models (cf.~\cref{subsec:MicroserviceArchitectureModelConstruction}). 
In the future, we plan to address this limitation by providing LEMMA technology models and code generators for languages targeting the \emph{Infrastructure as Code}~\cite{Morris2016} paradigm, e.g., Terraform~\cite{Brikman2019}. As a result, LEMMA would support model-based deployment to a variety of cloud-based deployment platforms.

In order to implement and take advantage of the LEMMA-based workflow, team members need to learn and use a new technology with LEMMA. As the validation (cf.~\cref{sec:validation}) shows, teams can significantly increase efficiency through the available generation facilities of LEMMA.  However, we need further empirical evaluation in practice (cf.~\cref{sec:conclusion}) to more accurately assess in which cases the efficiency gains from better documentation, accessible architectural understanding, and generation of deployment specifications outweigh the effort required to learn LEMMA and in which cases they do not.

An important aspect on which the efficiency of the workflow depends is the organization-wide agreement on the level of detail of the models shared between teams. For example, if a very high level of detail is agreed upon, i.e., including as much information as possible from the source code in the models, as we applied to the reconstruction of the case study (cf.~\cref{tab:LoCval}), generated artifacts must be more refined by the DevOps teams, i.e., a higher effort is necessary. This can be seen, for example, when looking at the Customer Core Service (cf.~\ref{sec:validation}).  The reconstructed model contains considerably more LoC, e.g. regarding technologies, than the generated model. In contrast, if the organization agrees on a low level of detail that, e.g., the abstraction from technology-related information and, thus, only considers technology-agnostic domain, service, and operation models (cf. LEMMA-Background-Subsection), very few adjustments to the generated models are necessary. 

A technical limitation within the LEMMA-based workflow is the unidirectional artifact creation. Changes to the models currently have to be made by the team owning the corresponding microservice. However, in order to further extend a shared understanding of the architecture as well as to follow DevOps' \emph{minimize communication efforts} characteristic~\cite{Ebert2016}, it would be beneficial if other teams or stakeholders could request editing of services of other teams directly using the shared models, e.g., to add an attribute to an interface operation. 

\section{Related Work}
\label{sec:related-work}
In the following, we describe related work from the areas of service and operation modeling, comparable qualitative studies, and workflows for DevOps-oriented development of microservice architectures in the context of model-driven software engineering.

\paragraph{MSA Service Modeling} Terzi{\'c} et al.~\cite{Terzic2018} present MicroBuilder, a tool that enables the modeling and generation of microservices. Like LEMMA, MicroDSL is based on the Eclipse Modeling Framework. Unlike LEMMA, however, MicroBuilder is closely linked to Java and Spring as specific technologies, so that the MicroDSL metamodel would have to be adapted for new technologies. MicroBuilder also addresses only the role of the developer and neglects stakeholders such as domain experts or operators. In addition, MicroBuilder does not address MSA's characteristic of having multiple teams involved in the development process. Another model-based approach called MicroART~\cite{Granchelli2017} is provided by Granchelli et al. MicroART contains a DSL called MicroARTDSL which aims to capture architecture information. The purpose of MicroART is to recover microservice architectures through static and dynamic analysis. As such, MicroART can support organizations in raising a common architectural understanding similar to the visualization we proposed in \cref{subsec:lemmaVisualizer}. However, MicroART does not provide a model-based workflow for the teams and lacks the rich ecosystem of LEMMA comprising means to also model and generate domain data, operational aspects, and different technologies.    

\paragraph{Qualitative Study} Bogner et al.~\cite{Bogner2019} describe a study related to our qualitative empirical analysis (cf.~\cref{sec:qualitative-study}) that includes 14 interviews with software architects. In contrast to our analysis, Bogner et al. do not focus on the challenges in the workflow of the organizations, but on the technologies used and software quality aspects. Another interview study was conducted by Haselb{\"o}ck et al.~\cite{Haselboeck2018} focusing on software design aspects such as the sizing of microservices. A questionnaire based study on Bad Smells in MSA was conducted by Taibi et al.~\cite{Taibi2018}. The study touches on organizational aspects and is included in our argumentation of the challenges (cf.~\cref{subsubsec:challenge1understanding}, but due to the study design as a questionnaire, the development process as a whole was not considered.

\paragraph{Development Workflows} In the context of our proposed workflow (cf.~\cref{subsec:workflow}), there are several \emph{large-scale agile} process models or methodologies that can foster the development of MSA by multiple DevOps teams.  Examples include Scrum at Scale~\cite{ScrumScale2020}, the Spotify Model~\cite{Smite2019}, or SAFe~\cite{SAFe2019} (cf.~\cref{subsec:MicroserviceDevProcess}). However, these approaches generally only become viable when an organization has at least 50 or more developers involved~\cite{Dikert2016}, and are therefore not suitable for SMOs facing the challenge of initially scaling their small organization from one to two or three teams. In addition, the aforementioned approaches address development at an organizational level and do not address development practices. Therefore, we expect our proposed workflow (cf.~\cref{subsec:workflow}) to integrate well with the stated large-scale approaches. 

\paragraph{MSA Operation Modeling}
The essential deployment metamodel (EDMM)~\cite{Wurster2020} is an approach that combines existential components of the deployment of a software system in a metamodel, taking into account concepts such as \emph{configuration management}~\cite{Delaet2010} and \emph{infrastructure as code}~\cite{Morris2016}. EDMM makes a specific mapping concerning the technology used for the software system's provisioning process based on the metamodel. For deploying the microservice application, EDMM supports technologies like Puppet\furl{https://puppet.com/}, Terraform\furl{https://www.terraform.io/}, AWS Cloud Formation\furl{https://aws.amazon.com/}, and Cloudify\furl{https://cloudify.co/}. Unlike EDMM, LEMMA addresses the deployment of service-based systems and their data structures and service composition. Besides, EDMM provides mapping concerning specific cloud providers. On the other hand, LEMMA provides technology-specific provisioning artifacts that can be used with different cloud providers. DICER~\cite{Artac2018} represents an approach based on technology-independent models for the generation of infrastructure as code and is used to deploy the software system. DICER models encapsulate monitoring, self-adaptation, configuration management, server deployment, and software system deployment. Also, DICER fosters the transformation of models into artifacts for service deployment using TOSCA\furl{https://cloudify.co/tosca/} and other technologies. The functional scope of DICER relates exclusively to the provisioning or operation of the software system. Furthermore, DICER does not support the modeling of data structures or service composition. Like LEMMA, DICER also provides technology-specific artifacts that can be used for the deployment process. Additionally, it also provides a graphical representation in the form of UML deployment diagrams, which LEMMA does not provide on an operational view.

\section{Conclusion and Future Work}
\label{sec:conclusion}
In this paper, we have identified two key challenge areas for SMOs through an empirical analysis of an interview study (cf.~\cref{sec:qualitative-study}). First, it is challenging for SMOs to develop and maintain a common understanding of architecture in an organization that is scaling to multiple teams for the first time through the application of SMA. Second, deployment in particular seems challenging due to its complexity, so SMOs tend to constitute special operation teams contrary to the microservice ownership principle (cf.~\cref{subsec:MicroserviceDevProcess}). This is detrimental to the implementation of DevOps practices and the benefits hoped for within the teams. 

To address these two challenge areas, we have presented a model-driven workflow based on LEMMA (cf.~\cref{sec:lemma}) for developing microservice architectures (cf.~\cref{sec:workflow}) and elaborated on the components we have added to LEMMA to support this workflow. The components comprise (i) interoperability bridges between OpenAPI and LEMMA models (cf.~\cref{subsec:apispecifications}); (ii) an extension to the Service Modeling Language to allow the import of remote models (cf.~\cref{subsec:assemblingSystemModel}); (iii) a model processor to visualize microservice architectures (cf.~\cref{subsec:lemmaVisualizer}); (iv) enhancing the Operation Modeling Language through the ability to import infrastructural nodes (cf.~\cref{subsec:InfrastructureModels}); and (v) code generators for microservice deployment and operation (cf.~\cref{subsec:InfrastructureCodeGeneration}).

For future work we plan to conduct a qualitative observation and interview study which aims to evaluate the proposed workflow in practice. Regarding the presented LEMMA extensions, we are going to mature the prototypical development and improve accessibility for users, e.g. by providing a dashboard. Furthermore, we would like to develop LEMMA's means to support a common architectural understanding in an organization not only through the presented visualization but also through analytical means such as code metrics.

\bibliographystyle{spmpsci}
\bibliography{literature}
\end{document}